\documentclass[12pt,reqno]{article}

\usepackage{amsmath, amsfonts, amssymb}
\allowdisplaybreaks
\numberwithin{equation}{section}

\renewcommand{\d}{\mathrm{d}}

\newcommand{\e}{\mathrm{e}}

\newcommand{\w}{\wedge}
\newcommand{\half}{\frac{1}{2}}
\DeclareSymbolFont{AMSa}{U}{msa}{m}{n}
\DeclareSymbolFont{AMSb}{U}{msb}{m}{n}
\DeclareMathSymbol{\fieldR}{\mathalpha}{AMSb}{"52}

\begin{document} 

\begin{flushright} \small

 ITP--UU--06/31 \\ SPIN--06/27 \\ hep-th/0607055

\end{flushright}

\bigskip
\begin{center}
 {\LARGE Supergravity description of spacetime instantons} \\[5mm]
Mathijs de Vroome and Stefan Vandoren\\[3mm]
 {\small\slshape
 Institute for Theoretical Physics \emph{and} Spinoza Institute \\
 Utrecht University, 3508 TD Utrecht, The Netherlands} \\
 {\upshape\ttfamily M.T.deVroome, vandoren@phys.uu.nl}
\end{center}

\vspace{10mm}
\hrule\bigskip

\centerline{\bfseries Abstract}\medskip

We present and discuss BPS instanton solutions that appear in type
II string theory compactifications on Calabi-Yau threefolds. From
an effective action point of view these arise as finite action
solutions of the Euclidean equations of motion in four-dimensional
$N=2$ supergravity coupled to tensor multiplets. As a solution
generating technique we make use of the c-map, which produces
instanton solutions from either Euclidean black holes or from
Taub-NUT like geometries.

\bigskip
\hrule\bigskip

\section{Introduction}

Black holes in superstring theory have both a macroscopic and
microscopic description. On the macroscopic side, they can be
described as solitonic solutions of the effective supergravity
Lagrangian. Microscopically they can typically be constructed by
wrapping $p$-branes over $p$-dimensional cycles in the manifold that
the string theory is compactified on. The microscopic interpretation
is best understood for BPS black holes.

Apart from this solitonic sector, string theory also contains
instantons. Microscopically they arise as wrapped Euclidean
$p$-branes over $p+1$-dimensional cycles of the internal manifold.
The aim of this paper is to present a macroscopic picture of these
instantons as solutions of the Euclidean equations of motion in
the effective supergravity Lagrangian. We focus hereby on
spacetime instantons, whose effects are inversely proportional to
the string coupling constant $g_s$. The models that we will study
are type II string theories compactified on a Calabi-Yau (CY)
threefold. The resulting effective action is $N=2, D=4$
supergravity coupled to vector multiplets and tensor multiplets.
The latter can be dualized to hypermultiplets, and the geometry of
the hypermultiplet moduli space - containing the dilaton - is
known to receive quantum corrections, both from string loops
\cite{RSV} and from instantons \cite{BBS}. The instanton
corrections are exponentially suppressed and are difficult to
compute directly in string theory. Our results yields some
progress in this direction, since within the supergravity
description one finds explicit formulae for the instanton
action.\footnote{Instanton actions can also be studied from
worldvolume theories of D-branes. For a discussion on this in the
context of our paper, we refer to \cite{MMMS}. It would be
interesting to find the precise relation to our analysis.} Related
work can also be found in \cite{GS2,BGLMM}, but our results are
somewhat different and contain several new extensions.

Interestingly, there is a relation between black hole solutions in
type IIA/B and instanton solutions in type IIB/A. Microscopically,
this can be understood from T-duality between IIA and IIB.
Macroscopically, this follows from the c-map \cite{CFG,FS}, as we
will show explicitly. This defines a map between vector and tensor
multiplets and as a consequence, (BPS) solutions of the vector
multiplet Lagrangian are mapped to (BPS) solutions of the tensor-
or hypermultiplet Lagrangian. We will use this mapping in
Euclidean spacetimes. Roughly speaking, there are two classes of
solutions on the vector multiplet sector: (Euclidean) black holes
and Taub-NUT like solutions. These map to D-brane instantons and
NS-fivebrane instantons respectively. The distinguishing feature
is that the corresponding instanton actions are inversely
proportional to $g_s$ or $g_s^2$ respectively. For both type of
instantons, we give the explicit solution and the precise value of
the instanton action.

The D-brane instantons are found to be the solutions to the
equations obtained from c-mapping the BPS equations of
\cite{LWKM}. Their analysis contains also $R^2$ interactions, but they can
be easily switched off. The BPS equations then obtained are
similar, but not identical to the equations derived in \cite{BLS}.
In the derivation and description of D-brane instantons we find it
convenient to make the symplectic structure of the theory and its
equations manifest. The NS-fivebrane instantons are derived in a
different way, not by using the c-map. This is because the BPS
solutions in Euclidean supergravity coupled to vector multiplets
are not fully classified. We therefore construct the NS-fivebrane
instantons by extending the Bogomol'nyi-bound-formulation of
\cite{TV1}.

Ultimately, we hope to get a better understanding of
non-perturbative string theory. In particular, it is expected that
instanton effects resolve conifold-like singularities in the
hypermultiplet moduli space of Calabi-Yau compactifications, see
e.g. \cite{OV}. These singularities are closely related - by the
c-map - to the conifold singularities in the vector multiplet moduli
space due to the appearance of massless black holes \cite{Strom}.
Moreover, in combination with the more recent relation between black
holes and topological strings \cite{OSV}, it would be interesting to
study if topological string theory captures some of the
non-perturbative structure of the hypermultiplet moduli space. For
some hints in this direction, see \cite{RVV}. Finally, we remark
that instantons play an important role in the stabilization of
moduli. For an example related to our discussion, we refer to
\cite{DSTV}.

This paper is organized as follows: In section \ref{five} we treat
NS-fivebrane instantons in the context of $N=1$ supergravity. We
use this simple setup to introduce various concepts, e.g. the
c-map, which we use in later sections. Section \ref{univ} is
devoted to a review of instanton solutions in the universal
hypermultiplet of $N=2$ supergravity and their relation to
gravitational solutions of pure $N=2$ supergravity. Then in
section \ref{gen} we consider instanton solutions to the theory
obtained from arbitrary CY compactification of type II
superstrings. Some technical details are provided in appendices at
the end of this paper, including a treatment of electric-magnetic
duality in tensor multiplet Lagrangians.

\section{NS-fivebrane instantons}\label{five}

In this section, we give the $N=1$ supergravity description of the
NS-fivebrane instanton. The main characteristic of this instanton
is that the instanton action is inversely proportional to the square of
the string coupling constant. In string theory, such instantons
appear when Euclidean NS-fivebranes wrap six-cycles in the internal space,
and therefore are completely localized in both space and (Euclidean) time.

It is well known that Euclidean
NS-fivebranes in string theory are T-dual to Taub-NUT or more
generally, ALF geometries \cite{OV} (see also \cite{Tong}). We
here re-derive these results from the perspective of
four-dimensional (super-) gravity in a way that allows us to
introduce the c-map conveniently.

\subsection{A Bogomol'nyi bound}

We start with a simple system of gravity coupled to a scalar and
tensor in four spacetime dimensions,
\begin{equation}\label{N=1action}
{\cal L}^{m}  = - \frac{1}{2\kappa^{2}} e R(e) + \frac{1}{2} |\d
\phi|^{2} + \frac{1}{2}{\rm e}^{2\phi} |H|^{2}\ ,
\end{equation}
with
\begin{equation}\label{tensor}
H=\d B\ .
\end{equation}
We use form notation for the matter fields; see Appendix
(\ref{form}) for our conventions.

This model appears as a sub-sector of $N=1$ low-energy effective
actions in which gravity is coupled to $N=1$ tensor multiplets.
In our case we have one tensor multiplet that consists of the
dilaton $\phi$ and the NS two-form $B$. In four dimensions, a
tensor can be dualized into a scalar, such that only chiral multiplets
appear. We will not do this dualization for reasons that become clear
below.

The instanton solution can be found by deriving a Bogomol'nyi
bound on the Euclidean Lagrangian \cite{Rey},
\begin{equation}\label{N=1BPSaction}
{\cal L}^e=\half |{\rm e}^\phi \ast H \mp {\rm e}^{\phi} \d
\e^{-\phi}|^2 \mp \d ({\rm e}^{\phi} H)\ .
\end{equation}
Here, we have left out the Einstein-Hilbert term. It is well known that this
term is not positive definite, preventing us to derive a Bogomol'nyi bound
including gravity. In most cases, our instanton solutions are purely in the
matter sector, and spacetime will be taken flat.
The Bogomol'nyi equation then is
\begin{equation}\label{N=1BPS}
\ast H = \pm \d {\rm e}^{-\phi}\  .
\end{equation}
This implies that ${\rm e}^{-\phi}$ should be a harmonic function.
The $\pm$ solutions refer to instantons or anti-instantons. Notice
that the surface term in \eqref{N=1BPSaction} is topological in the sense
that it is independent on the
spacetime metric. It is easy to check that the BPS configurations
\eqref{N=1BPS} have vanishing energy momentum tensor, so that the
Einstein equations are satisfied for any Ricci-flat metric.

One can now easily evaluate the instanton action on this solution.
The only contribution comes from the surface term in
\eqref{N=1BPSaction}. Defining the instanton charge as
\begin{equation}
\int_{S^3}\,H=Q\ ,
\end{equation}
with $H$ the three-form field strength, we find\footnote{In the
tensor multiplet formulation, the instanton action has no
imaginary theta-angle-like terms. They are produced after
dualizing the tensor into an axionic scalar, by properly taking
into account the constant mode of the axion. In the context of
NS-fivebrane instantons, this was explained e.g. in \cite{DTV}.}
\begin{equation}\label{NS5-action}
S_{{\rm inst}}=\frac{|Q|}{g_s^2}\ .
\end{equation}
Here we have assumed that there is only a contribution from
infinity, and not from a possible other boundary around the
location of the instanton. It is easy to see this when spacetime
is taken to be flat. In that case the single-centered solution for
the dilaton is
\begin{equation}
{\rm e}^{-\phi}={\rm e}^{-\phi_\infty}+\frac{|Q|}{4\pi^2 r^2}\ ,
\end{equation}
which is the standard harmonic function in flat space with the origin removed.
We have furthermore related the string coupling constant to the asymptotic
value of the dilaton by
\begin{equation}\label{g-string}
g_s\equiv {\rm e}^{-\phi_\infty/2}\ .
\end{equation}
In our notation, this is the standard convention.

\subsection{T-duality and the c-map}

We will now re-derive the results of the previous subsection using the c-map.
Though no new results, it will enable us to set the notation and to prepare
for more complicated situations discussed in the next sections.

To perform the c-map, we dimensionally
reduce the action \eqref{N=1action}
and assume that all the fields are independent of one
coordinate. This can most conveniently be done by first choosing
an upper triangular form of the vierbein, in coordinates
$(x^m,x^3\equiv \tau), m=0,1,2$,
\begin{equation}\label{vierbein}
e_\mu^a=\begin{pmatrix}{\rm e}^{-{\tilde \phi}/2}{\hat e}_m^i &
{\rm e}^{{\tilde \phi}/2}{\tilde B}_m \cr 0 & {\rm e}^{{\tilde
\phi}/2}\end{pmatrix}\ .
\end{equation}
The metric then takes the form
\begin{equation}\label{grav-metric}
{\rm d}s^2 =  {\rm e}^{{\tilde \phi}} ({\rm d}\tau + {\tilde
B})^{2} + {\rm e}^{-{\tilde \phi}}\hat{g}_{mn}{\rm d}x^{m}{\rm
d}x^{n}\ ,
\end{equation}
and we take ${\tilde \phi},{\tilde B}_m$ and ${\hat g}_{mn}$ to be
independent of $\tau$. For the moment, we take $\tau$ to be one of
the spatial coordinates, but at the end of this section, we will
apply our results to the case when $\tau$ is the Euclidean time.
In our example, the Wick rotation is straightforward on the
scalar-tensor sector.

We have that $e={\rm e}^{-{\tilde \phi}}{\hat e}$, and the scalar
curvature decomposes as (ignoring terms that lead to total
derivatives in the Lagrangian)
\begin{equation}
- e R(e)= - {\hat e} R (\hat e) + \frac{1}{2} |\d \tilde \phi|^2 +
\frac{1}{2}{\rm e}^{2{\tilde \phi}} |\tilde{H}|^2\ ,
\end{equation}
Similarly, we require the dilaton and 2-form to be independent of
$\tau$. The three-dimensional Lagrangian then is~\footnote{For
convenience of normalization, we set $\kappa^{-2}=2$.}
\begin{equation}\label{lns}
{\cal L}_3^m = - \hat{e} R(\hat e) + \frac{1}{2} |\d \tilde
\phi|^2 + \frac{1}{2}{\rm e}^{2{\tilde \phi}} |\tilde H|^2 +
\frac{1}{2} |\d \phi|^2 +\frac{1}{2}{\rm e}^{2{\phi}} |H|^2\ ,
\end{equation}
where $H$ is the two-form field strength descending from the
three-form $H$ in four dimensions (see also Appendix
(\ref{dimred})), so we have again that $H={\rm d}B$ in three
dimensions, where $B$ is a one-form.

In addition, there is an extra term in the Lagrangian,
\begin{equation}\label{L-aux}
{\cal L}_3^{aux}= - \frac{1}{2}{\rm e}^{2({\phi}+{\tilde \phi})}
|\mathbf{H}- \tilde{B} \wedge H|^2\ ,
\end{equation}
which plays no role in the three-dimensional theory. Here
$\mathbf{H} \equiv \d \mathbf{B}$ is the three-form arising from the
spatial component of the four-dimensional $H$. Being a three-form
in three dimensions it is an auxiliary field. This term can
therefore trivially be eliminated by its own field equation.

Note that the Lagrangian ${\cal L}_3$ has the symmetry
\begin{equation}\label{c:phi}
\phi \longleftrightarrow {\tilde \phi}\ ,\qquad B
\longleftrightarrow {\tilde B}\ .
\end{equation}
In fact, careful analysis shows that also ${\cal L}_3^{aux}$ is
invariant, provided we transform
\begin{equation}\label{c:B}
\mathbf{B} \rightarrow  \tilde{\mathbf{B}} \equiv \mathbf{B} -
\frac{1}{2} {\tilde B} \wedge B\ .
\end{equation}

The transformations in \eqref{c:phi} and \eqref{c:B} define the
c-map. The resulting theory can now be reinterpreted as a
dimensional reduction of a four-dimensional theory of gravity
coupled to a scalar $\tilde \phi $ and tensor ${\tilde B}$
obtained from the c-map, and vierbein
\begin{equation}\label{vierbein2}
{\tilde e_\mu^a}=\begin{pmatrix}{\rm e}^{-{\phi}/2}{\hat e}_m^i &
{\rm e}^{{\phi}/2}{B}_m \cr 0 & {\rm e}^{{\phi}/2}\end{pmatrix}\ ,
\end{equation}
where $\phi$ and $B$ are the original fields in \eqref{N=1action}
before the c-map. Our symmetry is related to the Buscher rules for T-duality
\cite{Bus}. We here derived these rules from an effective action
approach in Einstein frame, similar to \cite{T-dual}.

One can apply the c-map to solutions of the equations of motion. Given a
($\tau$-independent) solution $\{e_\mu^a,\phi, B_{\mu\nu}\}$, one
can construct another solution after the c-map, given by
$\{{\tilde e}_\mu^a,{\tilde \phi}, {\tilde B_{\mu\nu}}\}$ as
described above. This procedure can be done both in Minkowski and
in Euclidean space. In the latter case, we can take the coordinate
$\tau$ to be the Euclidean time, as time-independent solutions can
easily be Wick rotated. This is precisely the situation we are
interested in. To be more precise, we first formulate the Euclidean
four-dimensional theory based on Euclidean metrics coupled to a scalar
and tensor. The dimensional
reduction is still based on the decomposition of the vierbein
\eqref{vierbein} with $\tau$ the Euclidean time. After dimensional
reduction over $\tau$, the Einstein-Hilbert term now gives
\begin{equation}\label{Eucl-red}
e R(e)= {\hat e} R(\hat e) +\frac{1}{2} |\d \tilde{\phi}|^2
+\frac{1}{2}{\rm e}^{2{\tilde \phi}} |{\tilde H}|^2 ,
\end{equation}
such that the symmetry \eqref{c:phi} still holds.

\subsection{Taub-NUT geometries and NS-fivebrane instantons}

To generate instanton solutions, we will start from a time
independent solution of pure Einstein gravity, and perform the
c-map. This uplifts to a new solution in four dimensions with
generically nontrivial scalar and tensor. In other words, we do a
T-duality over Euclidean time. This of course only makes sense as
a solution-generating-technique. However, such a solution is not
an instanton, since it is not localized in $\tau$. We therefore
have to uplift the solution to a $\tau$-dependent solution in four
dimensions. This is easy if the original solution is in terms of harmonic
functions. In
that case there is a natural uplifting scheme, which involves
going from three- to four-dimensional harmonic functions.

We discuss now examples in the class of gravitational instantons
\cite{G-H}. These are vacuum solutions of the Euclidean Einstein
equation, based on a three-dimensional harmonic function $V({\vec
x})$,
\begin{equation}\label{grav-inst}
{\rm d}s^2=V^{-1}({\rm d}\tau + A)^2 + V {\rm d}{\vec x}\cdot
{\rm d}{\vec x}\ .
\end{equation}
Here $A$ is a one-form in three dimensions satisfying
$\ast \d A = \pm \d V$. The $\pm$ solutions yields
selfdual or anti-selfdual Riemann curvatures. In the notation of
\eqref{grav-metric}, we have that $A={\tilde B}$, ${\hat
e}_m^i=\delta_m^i$ and ${\rm e}^{-{\tilde \phi}}=V$.

Multi-centered gravitational instantons correspond to
harmonic functions of the form
\begin{equation}
V=V_0 + \sum_i \frac{m_i}{|\vec{x}-\vec{x}_i|}\ ,
\end{equation}
for some parameters $V_0$ and $m_i$. For non-zero $V_0$, one can
further rescale $\tau=V_0 {\tilde \tau}$ and $m_i=4V_0{\tilde m}_i$ such
that one can effectively set $V_0=1$. The single-centered case corresponds
to Taub-NUT geometries, or orbifolds thereof. For $V_0=0$ one obtains
smooth resolutions of ALE spaces (like e.g. the Eguchi-Hanson metric for the
two-centered solution). For more details, we refer to \cite{Andreas:1998hh}.

Before the c-map, the dilaton and two-form are taken to be zero. The
three-dimensional $\tau$-independent solution after the c-map is
\begin{equation}
{\rm e}^{\phi}=V^{-1}\ ,\qquad \ast H =\pm \d V\ , \qquad
\mathbf{H} =0\ ,
\end{equation}
and the metric is flat, $g_{mn}=\delta_{mn}$.

We now construct a four-dimensional $\tau$-dependent solution by
taking $V$ a harmonic function in four dimensions. We take the
four-dimensional metric to be flat and $H$ is still determined by
$\ast H=\pm {\rm d}V$, but now as a three-form field strength.

That this is still a solution for \eqref{N=1action}
can directly be seen from the fact that the Bogomol'nyi equations
\eqref{N=1BPS} are
satisfied. The instanton action is again given by
\eqref{NS5-action}. Notice further that the difference between
instantons and anti-instantons for the fivebrane corresponds to
selfdual and anti-selfdual gravitational instantons.

Due to our procedure, we are making certain aspects of T-duality
not explicit. We have for instance suppressed any dependence on the
radius of the compactified circle parameterized by $\tau$. These
aspects become important in order to dynamically realize the
uplifting solution in terms of a decompactification limit after
T-duality. It turns out that a proper T-duality of the Taub-NUT
geometry, including world-sheet instanton corrections, produces a
completely localized NS-fivebrane instanton based on the
four-dimensional harmonic function given above. For more details,
we refer to \cite{Tong}.

\section{Membrane and fivebrane instantons}\label{univ}

In the previous section, we have discussed aspects of NS-fivebrane
instantons. Here, we will elaborate further on this,
and also introduce membrane instantons. These appear in
M-theory or type IIA string theory compactifications, and
we will be interested in four-dimensional effective theories with
eight supercharges such as IIA strings compactified on Calabi-Yau manifolds.
The main distinction with the previous section is the presence of
RR fields, and these will play an important role in this section.

General CY compactifications of type IIA strings yield $N=2$
supergravity theories coupled to $h_{1,1}$ vector multiplets and
$h_{1,2}+1$ hypermultiplets (or tensor multiplets), but in this
section we will restrict ourself to the case of the universal
hypermultiplet only, leaving the general case for the next
section. This situation occurs when the CY space is rigid, i.e.
when $h_{1,2}=0$. Then there are only two three-cycles in the CY,
around which the Euclidean membranes can wrap. These are the
membrane instantons, and in this section we give their
supergravity description. The $h_{1,1}$ vector multiplet fields
can be truncated in our setup; it suffices to have pure
supergravity coupled to the universal hypermultiplet.

\subsection{Instantons in the double-tensor multiplet}

We will describe the universal hypermultiplet in the double-tensor
formulation, as this is what we get from the c-map. In this
formulation it contains two tensors and two scalars, which can be
thought of as two $N=1$ tensor multiplets coming from the NS-NS
and RR sectors.  Instantons in the double-tensor multiplet were
already discussed in \cite{TV1,DDVTV} (see also \cite{GS1}), and
in the context of the c-map in \cite{BGLMM}. In this section, we
reproduce these results and extend the c-map to include also
NS-fivebrane instantons.

The (Minkowskian) Lagrangian for the double-tensor multiplet can
be written as \cite{TV1,TV2}
 \begin{equation} \label{DTM-action}
  {\cal L}^m = - \d^4x\, e\, R + \half |F|^2+
\half |\d\phi|^2 + \half\,
  \e^{-\phi} |\d\chi|^2 + \half M_{ab} *\! H^a \w H^b \ .
 \end{equation}
The $N=2$ pure supergravity sector contains the metric and
the graviphoton field strength $F$, whereas the matter sector
contains two scalars and a doublet of 3-forms $H^a={\rm d}B^a$.
The self-interactions in the double-tensor multiplet are encoded
in the matrix
 \begin{equation}\label{M-matrix}
  M(\phi,\chi) = \e^{\phi} \begin{pmatrix} 1 & - \chi \\[2pt] - \chi &
\e^{\phi}  + \chi^2 \end{pmatrix}\ .
 \end{equation}
{}From a string theory point of view, the metric, $\phi$ and $H^2$
come from the NS sector, while the graviphoton, $\chi$ and $H^1$
descend from the RR sector in type IIA strings.
Notice that when we truncate to the NS
sector, we obtain the Lagrangian \eqref{N=1action}, so the results
obtained there are still valid here.

As we are interested in instanton solutions, we consider the
Euclidean version of (\ref{DTM-action}), which can be obtained by
doing a standard Wick rotation $t\rightarrow \tau=i t$ and using
Euclidean metrics. The form of the Lagrangian is still given by
(\ref{DTM-action}), but now the matter Lagrangian is positive definite.
In \cite{TV1} and \cite{DDVTV}, Bogomol'nyi equations were derived and solved
for the double-tensor multiplet coupled to pure $N=2$ supergravity with
vanishing graviphoton field strength and Ricci tensor. The solutions
of these equations preserving half of the supersymmetry can be recasted
into the following compact form:
\begin{eqnarray}\label{bpsinst}
g_{\mu \nu} & = & \delta_{\mu \nu}\ ,\nonumber\\
\e^{- \phi} & = & \frac{1}{4} (h^{2} - p^{2})\ ,\nonumber\\
\ast H^{2} & = & \frac{1}{2}(h \d p
- p \d h)\ ,\nonumber\\
\chi & = & - \e^{\phi}
p + \chi_{c}\ ,\nonumber\\
\ast (H^{1} - \chi_{c}H^{2}) & = & - \d h\ ,
\end{eqnarray}
with $h$ and $p$ four-dimensional harmonic functions (satisfying
$|h|\geq |p|$) and $\chi_c$ an arbitrary constant. The cases where
$h$ is negative or positive correspond to instantons or
anti-instantons respectively. We have written here a flat metric
$g_{\mu\nu}$, but it is easy to generalize this to any Ricci flat
metric, as long as it admits harmonic functions. Non-trivial $h$
and $p$ can be obtained when one or more points are taken out of
four-dimensional flat space. The solution is then of the form
\begin{equation}\label{h&p}
h=h_\infty + \frac{Q_h}{4 \pi^2 |\vec{x}-
\vec{x}_0|^{2}}\ ,\qquad p = p_\infty +
\frac{Q_p}{4 \pi^2 |\vec{x}-
\vec{x}_0|^{2}}\ ,
\end{equation}
or multi-centered versions thereof. It can be easily seen that a
pole in $p$ corresponds to a source with (electric) charge in the
field equation of $\chi$. Similarly a pole in $h$ corresponds to a
source with (magnetic) charge in the Bianchi identity of $H^1 -
\chi_c H^2$. For single-centered solutions, there are five
independent parameters, two for each harmonic function, together
with $\chi_c$.

\vspace{5mm}

{\bf NS-fivebrane instantons with RR background fields}

\vspace{5mm}

The general solution in \eqref{bpsinst} falls into two classes, depending
on the asymptotic behavior of the dilaton at the origin. The first
class fits into the category of NS-fivebrane instantons.  The solution
is characterized by
\begin{eqnarray}\label{pish}
p = \pm (h - \alpha)\ ,
\end{eqnarray}
with $\alpha$ an arbitrary constant. In terms of \eqref{h&p}, this
condition is equivalent to
\begin{equation}\label{QhisQp}
Q_h=\pm \,Q_p\ ,
\end{equation}
such that the solution only has four independent parameters.
This implies that the dilaton behaves at the origin like
\begin{equation}
\e^{-\phi} \rightarrow {\cal O}\left(\frac{1}{r^2}\right)\ .
\end{equation}
The condition \eqref{pish} implies that
\begin{equation}\label{BPS-RR-5brane}
\ast H^{2} = \pm \d \e^{-\phi}\ ,\qquad \ast H^{1} = \pm \d (\e^{-\phi}\chi)\ ,
\end{equation}
with $\e^{-\phi}$ the harmonic function
\begin{equation}\label{dil-RR-5brane}
\e^{-\phi} = \frac{1}{2} \alpha h - \frac{1}{4} \alpha^2\ ,
\end{equation}
and $\chi$ is fixed in terms of $h$ via \eqref{bpsinst}.
In this form, we get back the results of \cite{TV1} and \cite{DDVTV}.
The prototype example for $\e^{-\phi}$ is of the form
\begin{equation}
\e^{-\phi} = g_s^2 + \sum_i \frac{|Q_i|}{4 \pi^2 |\vec{x}-
\vec{x}_i|^{2}}\ .
\end{equation}
It is easy to check that, whereas the dilaton diverges,
the RR field $\chi$ remains finite at the excised points.
In fact, the BPS equations \eqref{bpsinst} require the values of
$\chi$ at the excised points ${\vec x}_i$ all to be equal \cite{DTV},
and we denote this value by $\chi_0$.

The solution is characterized by the parameters $\alpha,\chi_c,g_s$ and
the charges
\begin{equation}
Q \equiv \int_{S^3_\infty} H^2 = \mp \sum_i |Q_i|\  .
\end{equation}
The parameters $\alpha$ and $\chi_c$ can be traded for the boundary
values of the RR field $\chi_\infty$ and $\chi_0$.
The action of the multi-centered instanton was calculated in \cite{TV1,DDVTV},
and the result is
\begin{eqnarray}\label{dtns}
S_{\rm{inst}} = |Q| \Big( \frac{1}{g_{s}^{2}} + \frac{1}{2} (\Delta
\chi)^{2} \Big)\ ,
\end{eqnarray}
with $\Delta \chi \equiv \chi_\infty-\chi_0$.

The solution above describes a generalization of the NS-fivebrane
instanton discussed in section \ref{five}. Notice that the first term
in the instanton action is inversely proportional to the square of the string
coupling constant, as is common for NS-fivebrane instantons. The second
term is the contribution from the RR background field. Only for
constant $\chi$ does one obtain a local minimum of the
action~\footnote{Solutions with constant $\chi$ can be obtained from
\eqref{bpsinst} and \eqref{pish} by taking the limit $\alpha \rightarrow
0$ while both $h_\infty$ and $Q_h\rightarrow \infty$ in such a way that
$\alpha h$ is kept fixed. Such solutions follow more directly from the
Bogomol'nyi equations considered in \cite{DDVTV}.}.

\newpage

{\bf Membrane instantons}

\vspace{5mm}

The remaining solutions, other than \eqref{QhisQp}, are given by
\eqref{bpsinst} with $Q_h \neq Q_p$. One can see that the
asymptotic behavior of the dilaton around the origin is now
\begin{equation}
{\rm e}^{-\phi}\rightarrow {\cal O}\left(\frac{1}{r^4}\right)\ .
\end{equation}
Compared to the fivebrane instanton
case, this behavior is more singular. However, the instanton action is still
finite. As was shown in \cite{TV1}, the action reduces to a surface term, and
the only contribution comes from infinity. One way of writing the
instanton action is \cite{DDVTV}
\begin{equation}\label{meact}
S_{\rm inst}={\sqrt {\frac{4}{g_s^2}+(\Delta\chi)^2}}\left(|Q_h|
\pm \frac{1}{2}\Delta \chi \, Q\right)\ ,
\end{equation}
with the same convention as for fivebranes, i.e.
\begin{equation}\label{delch}
\Delta \chi \equiv \chi_\infty-\chi_0 = -\frac{p_\infty}{g_s^2}\ ,
\end{equation}
and $Q$ still defined by
\begin{equation}
Q \equiv \int_{S^3_\infty}\, H^2 = -\frac{1}{2}\left(h_\infty Q_p -
p_\infty Q_h\right)\ .
\end{equation}
The plus and minus sign in (\ref{meact}) refer to instanton and
anti-instanton respectively. Using the relations given above and
$g_s^2 = \frac{1}{4} (h_{\infty}^2 - p_{\infty}^2)$ one can show
that (\ref{meact}) is always positive, as it should be.

Notice that the instanton action contains both the fivebrane
charge $Q$ and $Q_h$, which we identify with a membrane charge.
For pure membrane instantons, which have vanishing NS-NS field,
the second term in (\ref{meact}) vanishes. When we put $H^2$ (and
its BPS equation) to zero from the start, we can dualize $\chi$ to
a tensor and obtain a "tensor-tensor" theory. To perform this
dualization we have to replace $\d \chi$ by the one-form $D$ in
the Euclideanized ($H^2$-less) version of (\ref{DTM-action}) and
add a Lagrange multiplier term
\begin{eqnarray} {\cal L}^e (\chi) \longrightarrow {\cal L}^e (D) + 2
i B_{\chi} \wedge \d D\ ,
\end{eqnarray}
where $B_{\chi}$ is a two-form. Integrating out $B_{\chi}$
enforces $\d D = 0$ and locally $D = \d \chi$ again. Subtracting
the total derivative $ 2 i \d (B_{\chi} \wedge  D)$ and
integrating out $D$ yields the tensor-tensor theory. Using this
action to evaluate the pure membrane instantons on gives
\begin{eqnarray}\label{meactt}
S'_{{\rm inst}} & = & S_{\rm inst} + \Delta \chi Q_p\nonumber\\
& = & \frac{2}{g_s} \sqrt{Q_h^2 - Q_p^2}\ .
\end{eqnarray}
The appearance of the second term in the first line is a result of
the subtraction of the boundary term in the dualization procedure.
In going from the first to the second line we used the fact that
$Q=0$, which allowed us to express $\Delta \chi$ in terms of the
charges $Q_h$ and $Q_p$ and $g_s$,
\begin{equation}
\Delta \chi = -\frac{2}{g_s}\frac{Q_p}{\sqrt {Q_h^2-Q_p^2}}\ .
\end{equation}
The $\frac{1}{g_s}$ dependence in the instanton action
is typical for D-brane instantons
that arise after wrapping Euclidean D-branes over supersymmetric
cycles in the Calabi-Yau\, \cite{BBS}.

The microscopic interpretation of the general solution is not so clear.
In the next subsection, we will see how these solutions are generated
from the c-map. In this way, one can give a natural interpretation in terms
of black holes and gravitational instantons.

The form of the instanton action for both fivebrane (\ref{dtns})
and membrane instantons  (\ref{meact}) and (\ref{meactt}) was
recently re-derived by solving the constraints from supersymmetry
of the effective action \cite{ASV}. This provides an alternative
derivation of the formulas in this section and confirms that the
supergravity method for computing the instanton action is correct.

\subsection{Einstein-Maxwell theory and the c-map}

In this subsection, we show that our membrane and fivebrane instanton
solutions naturally follow from the c-map. To show this, we start again
with the  Lagrangian \eqref{DTM-action} for Einstein-Maxwell theory
coupled to the double-tensor multiplet. After Wick rotating to Euclidean
space, we perform a dimensional reduction over
$\tau = it$. For the Minkowski theory, this was done in detail in
\cite{FS} and can be easily repeated for Euclidean signatures using
\eqref{Eucl-red}. We
decompose the metric as in \eqref{vierbein}; this yields a
three-dimensional metric, a one-form $\tilde B$ and a scalar
${\tilde \phi}$. The one-form gauge potential decomposes in the
standard way
\begin{equation}
A = (-{\tilde \chi}, {\tilde A} - {\tilde \chi}{\tilde B})\ ,
\end{equation}
where ${\tilde A}$ and ${\tilde B}$ are one-forms in three dimensions, and
${\tilde \chi}$ is the $\tau$ component of the four-dimensional gauge field.
More precisely, we have $A=-{\tilde \chi}{\rm d}\tau +
({\tilde A} - {\tilde \chi}{\tilde B})$.

The result after dimensional reduction is~\footnote{We are suppressing
here terms like \eqref{L-aux}, which are irrelevant for our purpose.}
 \begin{eqnarray} \label{3dDTM-action}
  {\cal L}_3^e &=& \d^3x\, {\hat e} R({\hat e}) + \half |\d\phi|^2 + \half\,
  \e^{-\phi} |\d\chi|^2 + \half M_{ab}(\phi,\chi) *\! H^a \w H^b \nonumber\\
&&\hspace{2.2cm}+ \half |\d{\tilde \phi}|^2 + \half\,
  \e^{-{\tilde\phi}} |\d{\tilde \chi}|^2 + \half M_{ab}({\tilde \phi},
{\tilde \chi}) *\! {\tilde H}^a \w {\tilde H}^b\ .
 \end{eqnarray}

Here we have combined the two one-forms in a doublet ${\tilde
B^a}=(\tilde{A},\tilde B)$ that define the (dual) two-form field strengths
${\tilde H}^a$ in three dimensions.
The matrix multiplying their kinetic energy is
exactly the same as in \eqref{M-matrix}, but now with the
tilde-fields. Therefore, the Lagrangian has the symmetry
\begin{equation}\label{tedu}
\phi  \longleftrightarrow  \tilde{\phi}\ ,\qquad \chi
\longleftrightarrow  \tilde{\chi}\ ,\qquad H^a \longleftrightarrow
{\tilde H}^a\ .
\end{equation}
Notice that, as observed before, setting the RR fields $\chi$ and
$H^1$ to zero reduces to the results of the previous section.

It is now clear that one can use the same technique as in the previous section,
namely to generate the instanton solutions from the solutions of the
Einstein-Maxwell Lagrangian, or vice versa. The symmetry transformations
(\ref{tedu}) basically interchange the gravitational and the double-tensor
multiplet degrees of freedom sectors of ({\ref{3dDTM-action}).

Since the general BPS instanton solution was given in \eqref{bpsinst}, it is
easy to translate this back to stationary BPS solutions of pure $N=2$
supergravity, after replacing four-dimensional by three-dimensional harmonic
functions. Starting from \eqref{bpsinst}, one obtains solutions to the
equations of motion in Euclidean space. Stationary solutions of the
Einstein-Maxwell Lagrangian can however easily be
continued from Minkowski to Euclidean space, and vice versa. If we make the
following decomposition for the metric and graviphoton vector field in
Minkowski space
\begin{eqnarray}\label{minmet}
g_{\mu \nu} \d x^{\mu} \d x^{\nu} & = & - \e^{\tilde{\phi}}(\d t +
\omega)^{2} +
\e^{-\tilde{\phi}} \hat{g}_{mn} \d x^{m} \d x^{n}\ ,\nonumber\\
A& = & (- \tilde{\chi}', \tilde{A} - \tilde{\chi}' \omega)\ ,
\end{eqnarray}
then we can analytically continue to Euclidean space by identifying:
\begin{equation}\label{thrwick}
\omega  =  -i {\tilde B}\ ,\qquad
\tilde{\chi}'  =  i \tilde{\chi}\ .
\end{equation}
BPS solutions of pure $N=2$ supergravity were studied
in \cite{GHU}, \cite{Tod} and \cite{BLS}. We are interested in stationary
solutions only, as non-stationary solutions cannot be used in the c-map.
Here and in the next section we will use the notation and results of the
analysis of \cite{LWKM}. In the case of pure $N=2$ supergravity the BPS
solutions, in terms of the variables of (\ref{minmet}), read:
\begin{eqnarray}\label{bpsgra}
\hat{g}_{mn} & = & \delta_{mn}\ ,\nonumber\\
\e^{- \tilde{\phi}} & = & \frac{1}{4} (h^{2} +
q^{2})\ ,\nonumber\\
\ast \d \omega & = & - \frac{1}{2} (h \d q
- q \d h)\ ,\nonumber\\
\tilde{\chi}' & = & - \e^{\tilde{\phi}}
q + \chi'_c\ ,\nonumber\\
\ast (\tilde{H}^{1} - \chi'_{c} \d \omega) & = & - \d h\ ,
\end{eqnarray}
with $h$ and $q$ three-dimensional flat space harmonic functions,
$\tilde{H}^1 = \d \tilde{A}$ and $\chi'_{c}$ an arbitrary
constant. The line element falls into the general class of
Israel-Wilson-Perj\'es (IWP) metrics \cite{IW,Perj},
\begin{equation}
{\rm d}s^2=-|U|^{-2} ({\rm d}t+\omega)^2+|U|^2 {\rm d}{\vec x}\cdot
{\rm d}{\vec x}\ ,
\end{equation}
where $U$ is any complex solution to the three dimensional Laplace equation.
Comparing to \eqref{bpsgra} and \eqref{minmet}, we have that
\begin{equation}
U=\frac{1}{2}(h+iq)\ .
\end{equation}

Let $F$ be the two-form field strength of the four-dimensional gauge field
and $G$ is its dual,
\begin{equation}\label{defig}
G_{\mu \nu} \equiv \frac{1}{2} \frac{\delta {\cal L}_4^m}{\delta F^{\mu
\nu}} = \frac{1}{2}(\ast F)_{\mu \nu}\ .
\end{equation}
Then we denote by $F_3$ and $G_3$ the corresponding three-dimensional
one-forms
\begin{equation}\label{drdcomp}
F_3  =  \d \tilde{\chi}'\ ,\qquad G_3  =  \frac{1}{2} \e^{\tilde
\phi} \ast(\omega \wedge F_3 - \mathbf{F})\ ,
\end{equation}
where $\mathbf{F}$ is the two-form arising from the spatial
components of the four-dimensional $F$. To derive the second
equation in (\ref{drdcomp}) one needs to decompose the component
of (\ref{defig}) with a time-index, $G_{tm} = \frac{1}{4}
\varepsilon_{tmnl} g^{n \mu} g^{l \nu} F_{\mu \nu}$, using the
metric parameterization (\ref{minmet}) (see Appendix
(\ref{dimred}) for our notations and conventions). The last two
equations in \eqref{bpsgra} can now elegantly be rewritten as:
\begin{equation}
F_{3}  =  - \d (\e^{\tilde{\phi}} q)\ ,\qquad
G_{3}  =  - \frac{1}{2} \d(\e^{\tilde{\phi}} h)\ .
\end{equation}
In fact in \cite{LWKM} solutions were given in terms of these objects. Note
that as $F$ and $G$ are electric-magnetic pairs, so are $F_3$ and
$G_3$. This will become important in the next section.

It is now clear that this set of BPS Einstein-Maxwell solutions
yields the same solutions as \eqref{bpsinst}, after analytic continuation
to the Euclidean fields given in (\ref{thrwick}), which amounts to
setting $p\equiv -i q$. The class of IWP metrics contains many interesting
examples, some of which we discuss now.

\vspace{5mm}

{\bf Pure membrane instantons and black holes}

\vspace{5mm}

We consider here solutions to (\ref{bpsinst}) with vanishing
NS-NS two-form:
\begin{equation}
H^{2} = 0\ .
\end{equation}
These were the solutions that lead to the pure membrane
instantons. The vanishing of $H^2$ implies that the two harmonic
functions $h$ and $p$ are proportional to each other,
\begin{equation}
p = c\, h\ ,
\end{equation}
for some real constant $c$. We take $h$ of the form:
\begin{equation}
h = h_{\infty} + \sum_i \frac{Q_{h,i}}{4 \pi^2 |\vec{x} -
\vec{x}_i|^2}\ , \qquad Q_{p,i}=c\,Q_{h,i}\ .
\end{equation}
This membrane instanton is in the image of the c-map. The dual
(Minkowskian) gravitational solution is static:
\begin{equation}
\d \omega = 0\ ,
\end{equation}
and has $q = c'\, h$. The IWP metric now becomes of the Majumdar-Papapetrou
type. These are multi-centered versions of the extreme
Reissner-Nordstr\"{o}m black hole. Our solutions describe the outer
horizon part of spacetime in isotropic coordinates,
\begin{eqnarray}\label{iso}
\d s^{2} = - \left(\gamma + \sum_i \frac{M_i}{4 \pi |\vec{x} -
\vec{x}_i| }\right)^{-2} \d t^{2} + \left(\gamma + \sum_i \frac{M_i}{4 \pi
|\vec{x} - \vec{x}_i|}\right)^{2} (\d r^{2} + r^{2} \d \Omega^{2})\ ,
\end{eqnarray}
with
\begin{equation}
M_i  =  \frac{1}{2} \sqrt{(Q_{h,i})^{2} + (Q_{q,i})^{2}}\ , \qquad
\gamma  =  \frac{1}{2}  \sqrt{1 + c'^2}\, h_{\infty}\ ,
\end{equation}
and $Q_{q,i} = c'\, Q_{h,i}$ for each charge labeled by $i$.
Note that in the parameterization
(\ref{iso}) the event horizons are located at $\vec{x} =
\vec{x}_i$. The metric can be made asymptotically Minkowski by a
rescaling of the coordinates
\begin{equation}
t  =  \gamma t'\ ,\qquad r  =  \frac{r'}{\gamma}\ .
\end{equation}

\vspace{5mm}

{\bf NS-fivebrane instantons and Taub-NUT with selfdual
graviphoton}

\vspace{5mm}

Here we consider the NS-fivebrane instantons with RR background fields.
This solution was specified by equations \eqref{pish}, \eqref{BPS-RR-5brane}
and \eqref{dil-RR-5brane}. Using the inverse
c-map, we can relate it to a BPS solution of pure N=2
supergravity, based on the three-dimensional harmonic function:
\begin{equation}
\e^{-\tilde{\phi}} = V \equiv v +  \sum_i \frac{Q_i}{4 \pi^2
|\vec{x}- \vec{x}_i|^{2}}\ .
\end{equation}
The metric solution of the Taub-NUT geometry (\ref{grav-inst}) then
reappears:
\begin{equation}\label{metr}
\d s^{2} = V^{-1} (\d \tau + \tilde{B})^{2} + V\, \d\vec{x}\cdot {\rm d}
{\vec x}\ ,
\end{equation}
with
\begin{equation}
\d \tilde{B} = \pm \ast \d V\ .
\end{equation}
Analogously to the NS-fivebrane instanton supporting a non-trivial
$\chi$, the Taub-NUT metric (\ref{metr}) supports a non-trivial
graviphoton:
\begin{eqnarray}\label{grvp}
F_3 = \pm \frac{1}{2} \alpha V^{-2} \d V\ , \qquad (\ast F)_3 = -
\frac{1}{2} \alpha V^{-2} \d V\ ,\nonumber\\
\mathbf{F} = \mp \frac{1}{2} \alpha \d \big( V^{-1} \tilde{B}
\big), \qquad \frac{1}{2} (\ast F)_{mn} \d x^m \wedge \d x^n  =
\frac{1}{2} \alpha \d \big( V^{-1} \tilde{B} \big)\ .
\end{eqnarray}
We remind that, in Euclidean space, the four-dimensional field strength is
given by $F = F_3 \wedge \d \tau + \frac{1}{2}
\mathbf{F}_{mn} {\rm d}x^m \wedge {\rm d}x^n$. For the solution
\eqref{grvp}, it is (anti-)selfdual. In fact, it
is precisely the one found in \cite{EH} (see equation (4.15) in that
reference).

The fact that the graviphoton is (anti-)selfdual implies that it
has vanishing energy-momentum, which is consistent with the fact
that the Taub-NUT solution is Ricci-flat. Taub-NUT solutions with
(anti-)selfdual graviphoton and their T-duality relation with
NS-fivebranes played an important role in a study of the partition
sum of the NS-fivebrane \cite{DVV}.

\section{Instantons in matter coupled $N=2$
supergravity}\label{gen}
In the last section we considered instantons in the double-tensor
multiplet coupled to $N=2$ supergravity. Now we are interested in
instanton solutions of the general four-dimensional low energy
effective action which type II superstrings compactified on a
Calabi-Yau give rise to. In the absence of fluxes, this yields
(ungauged) $N=2$ supergravity coupled to vector and tensor
multiplets (or their dual hypermultiplets). Following the spirit
of the previous section, we will generate the tensor multiplets
from the c-map on the gravitational and vector multiplet sector.
This yields a double-tensor multiplet and $h_{1,2}/h_{1,1}$ tensor
multiplets for type IIA/B string theories. These tensor multiplets
can be dualized further to hypermultiplets, but similarly to the
previous section, we will not carry out this dualization. This
turns out to be the most convenient way to describe instanton
solutions, i.e. they are naturally described in the tensor
multiplet formulation.

In this section we use the c-map, properly continued to Euclidean
space, to map the BPS equations for the vector multiplets as found
in \cite{LWKM} (with the $R^2$-interactions which are present in
there switched off) to instantonic BPS equations for the tensor
multiplet theory \footnote{For some earlier work on vector
multiplet BPS equations see \cite{BLS}, \cite{D} and references
therein.}.

The picture that emerges is that all BPS black hole solutions
have their corresponding instantonic description after the
(Euclidean) c-map. For a generic tensor multiplet theory these
solutions all carry some RR-charge, and the instanton action is
inversely proportional to the string coupling. There should also
be NS-fivebrane instantons whose action is proportional to
$\frac{1}{g_s^2}$. However it is not clear for a generic tensor
multiplet theory how to get these from the Euclidean c-map. Therefore we
derive them in a way independent of the c-map.

\subsection{The tensor multiplet theory}\label{ttmt}

We start by discussing the tensor multiplet Lagrangian obtained
after the c-map \cite{FS}. Details of the derivation can be found
in the next subsection, or e.g. in \cite{BGLMM}. The result is
$N=2$ supergravity coupled to a double-tensor multiplet and $n$
tensor multiplets, with $n=h_{1,1}$ or $h_{1,2}$ in IIA or IIB
respectively. The bosonic Lagrangian, in Minkowski space, reads
\begin{eqnarray}\label{minten}
\mathcal{L}^{m}_{T} & = & - \d^{4} x \, e R + \frac{1}{2} |\d
\phi|^{2} + \frac{1}{2} \e^{2 \phi} |H|^{2} + 2
\mathcal{M}_{IJ} \ast \d X^{I} \wedge \d \bar{X}^{J}\nonumber\\
& &  - \e^{- \phi}\, {\rm Im} \mathcal{N}_{IJ} \ast \d \chi^{I} \wedge \d
\chi^{J} - \e^{\phi}\, {\rm Im} \mathcal{N}_{IJ} \ast (H^{I} - \chi^{I} H)
\wedge (H^{J} - \chi^{J} H)\nonumber\\
& &  - 2\, {\rm Re} \mathcal{N}_{IJ} \d \chi^{I} \wedge (H^{J} - \chi^{J}
H)\ .
\end{eqnarray}
We have left out the vector multiplet sector including the
graviphoton as it is not relevant for our purposes. This sector
can be easily reinstalled. The NS-NS part of the bosonic sector of
the (double-)tensor multiplets consists of the dilaton, $\phi$,
the $2$-form $B$ ($H \equiv \d B$) and the complex scalars $X^{I}$
($I = 0, 1, ..., n$), which are subject to the condition
\begin{eqnarray}\label{conditions}
N_{IJ} \bar{X}^{I} X^{J} & = & -1\ .
\end{eqnarray}
This condition originates from the special K\"ahler geometry in
the vector multiplet sector before doing the c-map. The vector
multiplet sector is determined by a holomorphic prepotential
$F(X^I)$, homogeneous of second degree, with $N_{IJ} = -i (F_{IJ}
- \bar{F}_{IJ})$ and $F_{IJ} =
\partial_I \partial_J F(X)$. The constraint \eqref{conditions} arises naturally
as a gauge choice for dilatations in the superconformal calculus
\cite{deWit:1984pk,deWit:1984px},
and together with the U(1) gauge symmetry
\begin{equation}\label{ueen}
X^I \longrightarrow \e^{i \alpha} X^I\ ,
\end{equation}
one can effectively eliminate one complex degree of freedom.

The RR part of the bosonic sector of the (double)-tensor
multiplets is formed by the (real) scalars $\chi^I$ and the (real)
$2$-forms $B^I$ ($H^I \equiv \d B^I$). Altogether we have (after
fixing the dilatation and $U(1)$ gauge symmetries) $4n + 4$
on-shell bosonic degrees of freedom, which is indeed the
appropriate number for $n$ tensor multiplets and a double-tensor
multiplet.

The complex scalars $X^I$ parameterize a manifold with metric
\begin{eqnarray}\label{defm}
\mathcal{M}_{IJ} & \equiv & N_{IJ} - \frac{N_{IK}N_{JL}\bar{X}^{K}X^{L}}{N_{MN} X^{M} \bar{X}^{N}}\ .\nonumber\\
\end{eqnarray}
The matrix ${\rm Im}\mathcal{N}_{IJ}$ appearing in the quadratic terms of
(\ref{minten}) is determined by
\begin{eqnarray}\label{defn}
\bar{\mathcal{N}}_{IJ} & \equiv & F_{IJ} - i\frac{N_{KI}
\bar{X}^{K} N_{LJ} \bar{X}^{L}}{N_{MN} \bar{X}^{M} \bar{X}^{N}}\ .
\end{eqnarray}
One could impose the gauge choices on the matrices ${\mathcal M}$ and
${\mathcal N}$ to end up with the kinetic terms of the physical fields
only. In the region of the special K\"ahler manifold where the NS-NS scalars
have positive kinetic energy, one can show that also the RR scalars
have positive kinetic energy \cite{Cremmer:1984hj}.

The case of the double-tensor multiplet discussed in the previous
section can be obtained by setting $n=0$ and taking the
prepotential to be $F(X)=-\frac{i}{4}(X^0)^2$. This leads to
$N_{00}=-1, {\mathcal M}_{00}=0$ and ${\mathcal
N}_{00}=-\frac{i}{2}$, and it is easy to check that the Lagrangian
\eqref{minten} reduces to \eqref{DTM-action}.

To find instanton solutions, we first have to analytically continue
to Euclidean space. For a generic $N=2$ system with scalars and tensors,
this was discussed in \cite{DTV}. The standard rules for the Wick rotation
give (see Appendix (\ref{wic}) for our conventions),
\begin{eqnarray}\label{euclten}
\mathcal{L}^{e}_{T} & = & + \d^{4} x \, e R + \frac{1}{2} |\d
\phi|^{2} + \frac{1}{2} \e^{2 \phi} |H|^{2} + 2
\mathcal{M}_{IJ} \ast \d X^{I} \wedge \d \bar{X}^{J}\nonumber\\
& &  - \e^{- \phi} {\rm Im} \mathcal{N}_{IJ} \ast \d \chi^{I} \wedge \d
\chi^{J} - \e^{\phi} {\rm Im} \mathcal{N}_{IJ} \ast (H^{I} - \chi^{I} H)
\wedge (H^{J} - \chi^{J} H)\nonumber\\
& &  - 2 i {\rm Re} \mathcal{N}_{IJ} \d \chi^{I} \wedge (H^{J} -
\chi^{J} H)\ .
\end{eqnarray}
Notice that the last term becomes imaginary, similar to a theta-angle-like
term. It will therefore be difficult to find a Bogomol'nyi bound
on the action. We will return to the issue of a BPS bound in the last
subsection.
In fact, as we will see, we need to drop the reality conditions
on the fields, as not all solutions we discuss below respect these
reality conditions. For the moment, we will simply complexify all
the fields\footnote{For instance, this means that we treat $X^I$ and
${\bar X}^I$ as independent complex fields. The action then only depends
on $X$ and ${\bar X}$ in a holomorphic way.}, and discuss below
which instanton solutions respect which reality conditions.

It is convenient to rewrite
(\ref{euclten}) as
\begin{eqnarray}\label{clut}
\mathcal{L}^{e}_{T} & = & + \d^{4} x \, e R\nonumber\\
& & + \frac{1}{2} \e^{2 \phi}
(|H|^2 + |F_I \d \bar{Y}^I - Y^I \d \bar{F}_I + c.c.|^2)\nonumber\\
& & + d_I \wedge c^I - d^I \wedge c_I + D_I \wedge C^I + D^I
\wedge C_I\ .
\end{eqnarray}
Here and below, by $c.c.$ we mean taking the complex conjugate before
dropping the reality conditions, and then treating $X$ and ${\bar X}$
as independent complex fields.

The $Y^I$ are rescaled versions of the complex scalars $X^I$,
\begin{equation}\label{defy}
Y^I \equiv  \e^{- \frac{1}{2} \phi} \bar{h} X^I\ ,\qquad
{\bar Y}^I \equiv  \e^{- \frac{1}{2} \phi} h {\bar X}^I\ ,
\end{equation}
with $h$ an arbitrary (space-dependent) phase factor. Using
(\ref{defy}) the condition (\ref{conditions}) becomes an equation
for $\e^{-\phi}$ in terms of $Y^I$ and ${\bar Y}^I$,
\begin{eqnarray}\label{conde}
\e^{-\phi} & = & -i\Big(Y^I \bar{F}_I(\bar Y) - \bar{Y}^I F_I(Y)\Big)\ .
\end{eqnarray}

The three-forms ($c^I, c_I$) and the one-forms ($d^I, d_I$) belong
to the NS-NS sector. They are defined as
\begin{eqnarray}\label{def}
\left( \begin{array}c c^I\\ c_I \end{array} \right) \equiv \left(
\begin{array}c - i \ast \d (Y^I - \bar{Y}^I)\\ - i \ast \d (F_I -\bar{F}_I)\end{array}
\right)\ , \quad \left( \begin{array}c d^I\\ d_I \end{array}
\right) \equiv \left( \begin{array}c \d (\e^{\phi} (Y^I + \bar{Y}^I))\\
\d (\e^{\phi}(F_I + \bar{F}_I))\end{array} \right)\ .
\end{eqnarray}
In the RR sector we have the three-forms ($C^I, C_I$) and the
one-forms ($D^I, D_I$)
\begin{eqnarray}\label{deff}
\left( \begin{array}{c} C^I\\ C_I \end{array} \right) & \equiv & \left( \begin{array}{c} H^I - \chi^I H\\
i \e^{- \phi} {\rm Im} \mathcal{N}_{IJ} \ast \d \chi^J + {\rm Re}
\mathcal{N}_{IJ} (H^J - \chi^J H)\end{array} \right)\ ,\nonumber\\
\left( \begin{array}{c} D^I\\ D_I \end{array} \right) & \equiv &
\left( \begin{array}{c} - i \d \chi^I\\ - \e^{\phi} {\rm Im}
\mathcal{N}_{IJ} \ast (H^J - \chi^J H) - i {\rm Re} \mathcal{N}_{IJ} \d
\chi^J \end{array} \right)\ .
\end{eqnarray}
One can show that $C_I$ and $D_I$ are the functional derivatives
of the Lagrangian (\ref{clut}) with respect to $D^I$ and $C^I$
respectively
\begin{eqnarray}
(\ast C_I)_{\mu} =  \frac{1}{2} \frac{1}{\sqrt{|g|}} \frac{\delta
{\cal L}^e_{T}}{\delta D^{\mu I}}\ ,\qquad  (\ast D_I)_{\alpha \beta
\gamma} = - \frac{1}{2} \frac{6}{\sqrt{|g|}} \frac{\delta
{\cal L}^e_{T}}{C^{\alpha \beta \gamma I}}\ .
\end{eqnarray}
This means that the set of equations of motion and Bianchi
identities of the (tensor multiplet) RR fields can be formulated
as
\begin{eqnarray}\label{equat}
\d \left( \begin{array}{c} C^I\\ C_I \end{array} \right) =  - i
\left( \begin{array}{c} D^I\\ D_I \end{array} \right) \wedge H\ ,
\qquad \d \left( \begin{array}{c} D^I\\ D_I \end{array} \right) =
0\ ,
\end{eqnarray}
where the upper equations are Bianchi identities and the lower
ones are field equations. The set of equations (\ref{equat}) is
invariant under the electric-magnetic duality transformations
$Sp(2n +2, \mathbb{R})$ (see Appendix \ref{emduality}). So ($C^I,
C_I$) and ($D^I, D_I$) are symplectic vectors, as are ($Y^I, F_I$)
(its transformation induces the appropriate transformation of
$\mathcal{N}_{IJ}$) and so ($c^I, c_I$) and ($d^I, d_I$). To write
our theory in terms of these symplectic vectors is useful when we
consider the c-map and when deriving BPS equations through the
c-map. This is because the vector multiplet theory on the other
side of the c-map can be formulated in terms of symplectic vectors
as well. The symplectic vectors on both sides turn out to be
related in a rather simple way.

\subsection{The c-map}
Next we do a dimensional reduction of (\ref{clut}). We
parameterize the metric as
\begin{eqnarray}\label{eucmet}
g_{\mu \nu} \d x^{\mu} \d x^{\nu} & = & \e^{\tilde{\phi}}(\d \tau
+ \tilde{B})^{2} + \e^{-\tilde{\phi}} \hat{g}_{mn} \d x^{m} \d
x^{n}\ .
\end{eqnarray}
This way we obtain
\begin{eqnarray}\label{clutdrie}
\mathcal{L}^{e}_{T3} & = & + \d^{3} x \, \hat{e} R (\hat{e})\nonumber\\
& & + \frac{1}{2} \e^{2 \phi}
(|H|^2 + |F_I \d \bar{Y}^I - Y^I \d \bar{F}_I + c.c.|^2)\nonumber\\
& & + d_I \wedge c^I - d^I \wedge c_I + D_I \wedge C^I + D^I
\wedge C_I\ .
\end{eqnarray}
Here we have suppressed the term $\half |\d \tilde{\phi}|^2 +
\half \e^{2 \tilde{\phi}} |\tilde{H}|^2$ and terms of the form (\ref{L-aux})
as they play no role in our discussion. ($d^I, d_I$) and ($D^I, D_I$) are
still one-forms,
while ($c^I, c_I$)and ($C^I, C_I$) are now two-forms. They are
again given by (\ref{def}) and (\ref{deff}), but now with $B$ and
$B^I$ being one-forms.

As said before we can also obtain (\ref{clutdrie}) from a
dimensional reduction of a theory of $n$ vector multiplets and
coupled to $N=2$ supergravity. The bosonic sector of this
four-dimensional theory is given by
\begin{eqnarray}\label{minvec}
{\cal L}_{V}^{m} & = & - \d^{4}x \, \tilde{e} R (\tilde{e}) + 2
\mathcal{M}_{IJ} \ast \d X^{I} \wedge \d \bar{X}^{J} + F^I \wedge
G_I\ ,
\end{eqnarray}
with
\begin{eqnarray}
(\ast G_I)_{\mu \nu} \equiv \frac{1}{\sqrt{|g|}} \frac{\delta
{\cal L}_V^m}{\delta F^{\mu \nu I}} = (- {\rm Im} \mathcal{N}_{IJ} \ast
F^J + {\rm Re} \mathcal{N}_{IJ} F^J)_{\mu \nu}\ .
\end{eqnarray}
We have left out the  - irrelevant - tensor multiplet sector, just
as we did with the vector multiplet sector in (\ref{minten}). The
bosonic NS-NS sector of (\ref{minvec}) consists of the metric
$\tilde{g}_{\mu \nu}$ and the complex scalars $X^I$ (again subject
to (\ref{conditions}) and a $U(1)$ gauge fixing condition). The
bosonic RR sector is formed by the one-forms $A^I$ ($F^I \equiv \d
A^I$). As is well known, the set of field equations and Bianchi
identities of $F^I$ is invariant under $Sp(2n+2, \mathbb{R})$,
i.e. ($F^I, G_I$) is a symplectic vector.

Now we do a dimensional reduction over time, using
\begin{eqnarray}\label{minmett}
\tilde{g}_{\mu \nu} \d x^{\mu} \d x^{\nu} & = & - \e^{\phi}(\d t +
\omega)^{2} +
\e^{-\phi} \hat{g}_{mn} \d x^{m} \d x^{n}\ ,\nonumber\\
A^I & = & (- \chi^{I'}, B^I - \chi^{I'} \omega)\ .
\end{eqnarray}
Then after identifying
\begin{equation}\label{thrwickk}
\omega  =  -i B\ ,\qquad \chi^{I'}  =  i \chi^I\ ,
\end{equation}
complexifying all fields and multiplying the resulting
three-dimensional Lagrangian by $-1$, we re-obtain
(\ref{clutdrie}). As we already hinted at at the end of the last
subsection, we have the following simple relations between the
(RR) symplectic vectors coming from both sides
\begin{eqnarray}\label{relsym}
\left( \begin{array}c F_3^I\\ G_{3I} \end{array} \right) = -
\left( \begin{array}c D^I\\ D_I\end{array} \right)\ , \quad \left(
\begin{array}c \mathbf{F}^I\\ \mathbf{G}_I\end{array}
\right) = \left( \begin{array}c C^I\\ C_I\end{array} \right) - i
\left( \begin{array}c D^I\\ D_I\end{array} \right) \wedge B\ .
\end{eqnarray}

\subsection{BPS equations from the c-map}\label{equations}
In this section we use the c-map to obtain BPS instanton equations
of a $n+1$ tensor multiplet theory from the BPS equations of a $n$
vector multiplet theory. As said before, the latter equations are
known and we use the results of \cite{LWKM}. In here equations
were constructed for stationary solutions preserving half of the
supersymmetry with parameters satisfying
\begin{equation}
h \epsilon_{ij} = \varepsilon_{ij} \gamma_{0} \epsilon^{j}\ .
\end{equation}
We remind that $h$ is the phase factor appearing in (\ref{defy}).
The metric components were found to be related to the complex
scalars $Y^I$ in the following way
\begin{eqnarray}\label{bpslm4}
\e^{-\phi} & = & - i(Y^{J} \bar{F}_{J} - \bar{Y}^{J} F_{J})\ ,\nonumber\\
\hat{g}_{mn} & = & \delta_{mn}\ ,\nonumber\\
\ast \d \omega & = & - \bar{F}_{J} \d Y^{J} + \bar{Y}^{J} \d F_{J}
+ c.c.\ ,
\end{eqnarray}
while $-i(Y^I - \bar{Y}^I)$ and $-i(F_I - \bar{F}_I)$ should be
three-dimensional harmonic functions. This fixes the NS-NS sector
completely. Recall that the equation for $\e^{-\phi}$ is
identically true, as follows from the definition of $Y^I$  and the
condition (\ref{conditions}). For the RR fields the BPS equations
of \cite{LWKM} are
\begin{eqnarray}\label{RReq}
\left( \begin{array}c F_3^I\\ G_{3I} \end{array} \right) = -
\left( \begin{array}c d^I\\ d_I\end{array} \right)\ .
\end{eqnarray}
These equations can be equivalently formulated as
\begin{eqnarray}\label{RReqa}
\left( \begin{array}c \mathbf{F}^I\\ \mathbf{G}_I \end{array}
\right) = - \left( \begin{array}c c^I \\ c_I\end{array} \right) -
i \d \left( \begin{array}c \e^{\phi} (Y^I + \bar{Y}^I) B|_{\rm
BPS}\\ \e^{\phi} (F_I + \bar{F}_I) B|_{\rm BPS}\end{array}
\right)\ ,
\end{eqnarray}
where $B|_{\rm BPS}$ is the BPS solution of $B$. We remind that
$(c^I, c_I)$ and $(d^I, d_I)$ are given by (\ref{def}).
(\ref{RReqa}) is the form in which the equations for the RR fields
in \cite{BLS} are written, however they do not have the second
term on the r.h.s. Both sets of equations (\ref{RReq}) and
(\ref{RReqa}) fix the RR fields completely in terms of the complex
scalars $Y^I$.

By construction the equations above only have stationary
solutions. When $\omega = 0$ one gets static extremal black holes.
This works similar as in the pure supergravity case discussed in
the last section. However there is a difference between the
generic case and pure supergravity, which will become important
later on in the context of NS-fivebrane instantons. We saw in last
section that the pure $N=2$ supergravity BPS equations, after an
analytic continuation to Euclidean space, gave rise to Taub-NUT
solutions as well. In contrast to this, for generic functions
$F(X)$ it is far from clear if, and if yes how, this kind of
solutions is contained in the general solution.

The equations above can be mapped quite easily to instanton
equations of our Euclidean tensor sector. The first step is the
dimensional reduction over time. As these equations only have
stationary solutions, they are still valid after this dimensional
reduction. Of course then they are also valid equations of the
dimensionally reduced version of the tensor multiplet theory
(\ref{clutdrie}). The last step is to uplift them to equations of
the full tensor theory (\ref{clut}). Basically the equations
remain the same except that two-forms become three-forms and
three-dimensional harmonic functions become four-dimensional
harmonic functions.
We find as instanton equations for the NS-NS fields
\begin{eqnarray}\label{bpsensns}
\e^{-\phi} & = & - i(Y^{J} \bar{F}_{J} - \bar{Y}^{J} F_{J})\ ,\nonumber\\
g_{\mu \nu} & = & \delta_{\mu \nu}\ ,\nonumber\\
\ast H & = & - i(\bar{F}_{J} \d Y^{J} - \bar{Y}^{J} \d F_{J} +
c.c.)\ ,
\end{eqnarray}
while $-i(Y^I - \bar{Y}^I)$ and $-i(F_I - \bar{F}_I)$ are now
four-dimensional harmonic functions.

Using (\ref{relsym}) we directly read of what the instanton
equations for the RR fields are
\begin{eqnarray}\label{RReqb}
\left( \begin{array}c D^I\\ D_{I} \end{array} \right) =  \left(
\begin{array}c d^I\\ d_I\end{array} \right)\ ,
\end{eqnarray}
or
\begin{eqnarray}\label{RReqaa}
\left( \begin{array}c C^I\\ C_I\end{array} \right) =  \left(
\begin{array}c c^I\\ c_I\end{array}
\right) - i \left( \begin{array}c \e^{\phi} (Y^I + \bar{Y}^I)\\
\e^{\phi} (F_I + \bar{F}_I)\end{array} \right) H|_{inst} \ .
\end{eqnarray}
Just as on the vector multiplet side both (\ref{RReqb}) and
(\ref{RReqaa}) fix the RR fields completely in terms of the
complex scalars $Y^I$. For the fields appearing in (\ref{euclten})
the equations take the form
\begin{eqnarray}\label{BPS-chiI}
\chi^I & = & i \e^{\phi} (Y^I + \bar{Y}^I) + \chi^I_c\ ,\nonumber\\
\ast (H^I - \chi^I_c H) & = &  i  \d (Y^I - \bar{Y}^I)\ ,
\end{eqnarray}
where $\chi^I_c$ are arbitrary constants.

Recall from subsection (\ref{ttmt}) that all fields are complex.
However, when we take $-i(Y^I-{\bar Y}^I)$ and $-i(F_I-{\bar F}_I)$
to be real, then the solutions for the dilaton and $H^I$ are real
whereas $\chi^I$ and $H$ become imaginary.

\vspace{3mm}

Let us make contact with the results of section \ref{univ}. When
we take the function $F(Y)$ to be $F(Y) = - \frac{1}{4} i
(Y^0)^2$, we see that (\ref{minten}) reduces to the action
(\ref{DTM-action}). We then get
\begin{equation}\label{dteen}
Y^0 + \bar{Y}^0  =  2 i (F_0 - \bar{F}_0)\ ,\qquad
F_0 + \bar{F}_0  =  - \half i (Y^0 -\bar{Y}^0)\ .
\end{equation}
Now we make the following identification of the harmonic functions
$-i(Y^0 - \bar{Y}^0)$ and $-i(F_0 - \bar{F}_0)$ and the harmonic
functions $h$ and $p$ which appeared in the BPS equations of
section \ref{univ}
\begin{equation}\label{dttwee}
-i(Y^0 - \bar{Y}^0)  =  h\ ,\qquad
-i(F_0 - \bar{F}_0)  =  - \half i p\ .
\end{equation}
Equations (\ref{bpsinst}) then follow directly. We can in this case
obtain a real solution for $\chi$ and $H$ if we impose that
$-i(F_0-{\bar F}_0)$ is imaginary, such that the harmonic function $p$
is real.

\subsection{D-brane instantons}

We now discuss the different types of solutions to the equations
we obtained above. Clearly the general solution is a function of
$2n+2$ harmonic functions. In the following we take them
single-centered
\begin{eqnarray}\label{hp}
- i(Y^I - \bar{Y}^I) & = & - i(Y^I - \bar{Y}^I)_\infty +
\frac{\hat{Q}^I}{4 \pi^2 |\vec{x}- \vec{x}_0|^{2}}\ ,\nonumber\\
- i(F_I - \bar{F}_I) & = & - i(F_I - \bar{F}_I)_{\infty} +
\frac{Q_I}{4 \pi^2 |\vec{x}- \vec{x}_0|^{2}}\ .
\end{eqnarray}
However our results are easily generalized to multi-centered
versions of (\ref{hp}).

In section  \ref{univ} we saw that the two different types of
solutions to the BPS equations for the double-tensor multiplet,
membrane and the NS-fivebrane instantons, have different behavior
of $\e^{-\phi}$. For membrane instantons (having non-zero
RR-charge) the dilaton behaves towards the excised point(s) as
${\rm e}^{-\phi}\rightarrow {\cal O}\left(\frac{1}{|\vec{x} -
\vec{x}_0|^4}\right)$. For NS-fivebrane instantons (having
non-zero NS-NS-, but vanishing RR-charge) $\e^{-\phi}$ is a
harmonic function, which implies that towards the excised point(s)
the behavior of the dilaton is ${\rm e}^{-\phi}\rightarrow {\cal
O}\left(\frac{1}{|\vec{x} - \vec{x}_0|^2}\right)$. The different
behavior of the dilaton in both types of solutions is reflected in
a different dependence of the instanton action on the string
coupling.

Let us now consider solutions to the general equations of last
subsection (i.e. for general functions $F(Y)$). The above seems to
indicate that for a study of the characteristics of the instanton
solutions it is good to start by analyzing the behavior of the
dilaton towards the excised point(s). Doing this analysis we find
that to leading order in $\frac{1}{|\vec{x} - \vec{x}_0|}$
\begin{eqnarray}\label{dilor}
\e^{-\phi}|_{\vec{x} \rightarrow \vec{x}_0}  & = & \frac{|Z_0|^2}
{16 \pi^4 |\vec{x} - \vec{x}_0|^4}\ ,
\end{eqnarray}
which is as singular as the membrane instanton of section
\ref{univ}. Here $Z_0$ is defined as
\begin{equation}
Z_0 \equiv (\hat{Q}^I F_I (X) - Q_I X^I)|_{\vec{x} \rightarrow
\vec{x}_0}\ .
\end{equation}
As seen from the c-map, the function $Z = Q^I F_I (X) - Q_I X^I$
is the dual of the central charge function of the vector multiplet
theory.

\vspace{5mm}

$\mathbf{Z_0 \neq 0}$

\vspace{5mm}

We first consider the case $Z_0 \neq 0$. Generic single-centered
solutions consist of $5n +5$ parameters, $2$ for each harmonic
function and the $n+1$ constants $\chi^I_c$. The RR scalars
$\chi^I$ take the values $\chi^I_c$ at $\vec{x} = \vec{x}_0$. The
constants $\hat{Q}^I$ and $Q_I$ appearing in (\ref{hp}) can be
identified with magnetic and electric charges of sources appearing
in Bianchi identities and field equations respectively.
$\hat{Q}^I$ is equal to the charge of the source in the Bianchi
identity of $\hat{H}^I$
\begin{equation}
\hat{Q}^I = \int_{S^3_{\infty}} \hat{H}^I\ ,
\end{equation}
where we have defined $\hat{H}^I \equiv H^I - \chi^I_c H$. $Q_I$
is up to a factor $2i$ the charge of the source in the field
equation of $\chi^I$
\begin{equation}
2i Q_I = \int_{\mathbb{R}^4} \d^4 x \, (\frac{\delta {\cal L}}{\delta
\chi^I} -
\partial_{\mu} \frac{\delta {\cal L}}{\delta \partial_{\mu} \chi^I})\ .
\end{equation}
Notice that this is consistent with the fact that the solutions
for $\chi^I$ are imaginary. As there are non-vanishing RR charges
we can identify these solutions as D-brane instantons,
generalizing the membrane instantons found in section \ref{univ}.
Also the Bianchi identity of $H$ is sourced. The corresponding
charge can be expressed in terms of the parameters appearing in
the $2n+2$ harmonic functions
\begin{equation}
Q \equiv \int_{S^3_{\infty}} H = (F_I - \bar{F}_I)_{\infty}
\hat{Q}^I - (Y^I - \bar{Y}^I)_{\infty} Q_I\ .
\end{equation}

Evaluating (\ref{clut}) on these instantons gives
\begin{eqnarray}\label{tenact}
S^T_{\rm inst} & = & \int_{\mathbb{R}^4} (2 d_I \wedge c^I - i
({\rm e}^{\phi} (Y^I + \bar{Y}^I) d_I + {\rm e}^{\phi} (F_I (Y)
+ \bar{F}_I (\bar{Y})) d^I) \wedge H)|_{\rm BPS} \nonumber\\
& = & \frac{2}{g_s^2} (F_{I}(Y) + \bar{F}_{I}(\bar{Y}))_{\infty}
\hat{Q}^I - \frac{i}{g_s^4} (F_{I}(Y) +
\bar{F}_{I}(\bar{Y}))_{\infty} (Y^I + \bar{Y}^I)_{\infty} Q\ .
\end{eqnarray}

Applying (\ref{tenact}) to the double-tensor multiplet theory of
section \ref{univ}, we have to take again $F(Y) = - \frac{1}{4} i
(Y^0)^2$. Then using (\ref{dteen}), (\ref{dttwee}), (\ref{delch})
and the double-tensor multiplet relation $g_s^2 = \frac{1}{4}
(h_{\infty}^2 - p_{\infty}^2)$ we re-obtain (\ref{meact}).

For $Q=0$ the second term in (\ref{tenact}) vanishes and we find
\begin{eqnarray}\label{tenactt}
S^T_{\rm inst} & = & \frac{2}{g_s} (\bar{h}F_{I}(X) + h
\bar{F}_{I} (\bar{X}))_{\infty} \hat{Q}^I\ .
\end{eqnarray}
This is the action for pure D-brane instantons of which the pure
membrane instanton of section \ref{univ} is a specific example. We
have reintroduced the variables $X^I$ to make explicit the typical
$\frac{1}{g_s}$ dependence of D-brane instanton actions. From the
c-map point of view pure D-brane instantons are the duals of
static BPS black holes living in the vector multiplet sector.
Microscopically D-brane instantons come from wrapping even/odd
branes over odd/even cycles in the Calabi-Yau in type IIA/B string
theory.

Defining $\Delta\varphi_I \equiv \frac{i}{g_s} (\bar{h}F_{I}(X) + h
\bar{F}_{I} (\bar{X}))|_\infty$, we can rewrite
(\ref{tenactt}) as
\begin{equation}
S^T_{\rm inst} =  -2i \Delta \varphi_I Q^I\ .
\end{equation}
In fact, one can show that $\Delta\varphi_I=\varphi_{I\infty}-\varphi_{I0}$,
where $\varphi_I$ is the dual (RR) scalar of $\hat{H}^I$ and 
$\varphi_{I\infty}$ and $\varphi_{I0}$ are the asymptotic values of 
$\varphi_I$ evaluated on the BPS solution
\begin{equation}
\varphi_I=i{\rm e}^\phi(F_I(Y)+{\bar F}_I(\bar Y))+\varphi_{Ic}\ .
\end{equation}
Here $\varphi_{Ic}$ are integration constants, which coincide with 
$\varphi_{I0}$, the value of $\varphi_I$ at the point $\vec{x}=\vec{x}_0$.
This BPS equation is in fact implicitly stated already in the bottom equation
in (\ref{RReqb}). Observe furthermore that the BPS solutions for $\chi^I$, as
in (\ref{BPS-chiI}), and $\varphi_I$ are consistent with symplectic 
transformations, so we can write
\begin{eqnarray}\label{chi-phi}
\left( \begin{array}c \chi^I\\ \varphi_I\end{array} \right) =  
i \e^\phi \left( \begin{array}c Y^I + \bar{Y}^I\\
F_I + \bar{F}_I\end{array} \right) +
\left( \begin{array}c \chi^I_c\\ \varphi_{Ic}\end{array} \right)
\ .
\end{eqnarray}

Like in section \ref{univ}, when we
put $H$ (and its BPS equation) to zero from the start, we can
dualize all RR-scalars to tensors. This way we obtain a
formulation of the theory consisting of $2n+2$ tensors, the
"tensor-tensor" theory. The dualization procedure works similar as
the one described in section (\ref{univ}). First we write $D^I$
(being a one-form) instead of $\d \chi^I$ in the Euclideanized
version of (\ref{euclten}) (without $H$) and add a Lagrange
multiplier term
\begin{eqnarray}
{\cal L}^e_{T} (\chi^I) \longrightarrow {\cal L}^e_{T} (D^I) + 2 i B_I
\wedge \d D^I\ ,
\end{eqnarray}
where $B_I$ are $n+1$ two-forms. Integrating out $B_I$ enforces
$\d D^I = 0$, giving back (locally) $D^I = \d \chi^I$. Subtracting
the total derivative $ 2 i \d (B_I \wedge  D^I)$ and integrating
out $D^I$ yields the tensor-tensor theory. When we evaluate this
action on the pure D-brane instantons we get
\begin{eqnarray}\label{acttt}
S^{TT}_{{\rm inst}} & = & - 2 i \Delta \varphi_I Q^I  + 2 i\Delta \chi^I Q_I\nonumber\\
& = & \frac{2}{g_s} (\bar{h} F_{I}(X) + h \bar{F}_{I}
(\bar{X}))_{\infty} \hat{Q}^I - \frac{2}{g_s} (\bar{h} X^I + h
\bar{X}^I)_{\infty} Q_I\nonumber\\
& = & \frac{4}{g_s} |Z|_{\infty}\ .
\end{eqnarray}
The second term in the first line is due to the subtraction of the
boundary term in the dualization procedure. To arrive at the last
line we have used that $Q = 0$, which is consequence of the fact
that we have put $H$ to zero. The expression in the last line is
(up to a factor of $4$) the value of the real part of the pure
D-brane instanton action as suggested in (a five-dimensional
context) in \cite{GS2}.

\vspace{5mm}

$\mathbf{Z_0 = 0}$

\vspace{5mm}

The case $Z_0=0$ is special and needs to be analyzed separately,
as the behavior of the dilaton is different. For the double-tensor
multiplet of section \ref{univ} it yields NS-fivebrane instantons,
which have a harmonic $\e^{-\phi}$. This can most easily be seen
from the fact that in the double-tensor multiplet case we have
$|Z_0| = \frac{1}{2} \sqrt{Q_h^2 - Q_p^2}$ (for single-centered
instantons, as can be derived using (\ref{defy}), (\ref{dteen})
and (\ref{dttwee})). Requiring $|Z_0|$ to vanish then gives the
NS-fivebrane relation (\ref{QhisQp}). However, for generic
functions $F(X)$ things work differently and we do not get
NS-fivebrane instantons from taking $Z_0=0$. In fact in these
cases the $Z_0=0$ solution only differs qualitatively from the
$Z_0 \neq 0$ solution close to the excised points, which is
directly related to the fact that only the asymptotic behavior of
the dilaton is different. Now recall that $Z$ is the dual of the
central charge function of the vector multiplet theory. So $Z_0=0$
solutions are the duals of vector multiplet solutions with
vanishing central charge function at $\vec{x} = \vec{x}_0$. In
case $Q=0$ these are zero-horizon black holes. Just as higher
derivative corrections lift zero-horizon black holes at the
two-derivative-level to finite horizon black holes \cite{DKM}, we
expect that for $Z_0=0$-instantons higher derivative corrections
have a qualitative effect on the behavior of $\e^{-\phi}$ in the
limit $\vec{x} \rightarrow \vec{x}_0$. If this is the case the
(two-derivative) differences between these solutions and the $Z
\neq 0$ instanton have no real physical significance.

\subsection{NS-fivebrane instantons}

In the previous section, we have discussed D-brane instantons.
These were obtained from the c-map of the BPS solutions of [45],
analytically continued to Euclidean space. We also saw that for
generic functions $F(X)$ NS-fivebrane instantons did not appear as
a limiting case in a similar way as in the double-tensor multiplet
theory of section \ref{univ}. In fact it is not clear if they are
contained at all in the general solution to the equations in
subsection \ref{equations}, just as was the case for their
supposedly dual Taub-NUT solutions on the vector multiplet side.

However, we expect there to be (BPS) NS-fivebrane instantons in
the general theory as well. That we have missed them so far could
be understood from the fact that not all solutions in the
Euclidean theory can be obtained from Wick rotating real solutions
in the Lorentzian theory.  Therefore, we will follow a different
strategy and work directly in the Euclidean tensor multiplet
Lagrangian, using a similar method as in \cite{TV1}. This way we
indeed find a class of NS-fivebrane instanton solutions.

We first write (\ref{euclten}) as
\begin{eqnarray}\label{bbound}
{\cal L}^e_{T} & = & \d^{4} x \, e R + 2 \mathcal{M}_{IJ} \ast \d X^I \wedge \d \bar{X}^J\nonumber\\
& & + \ast (N \ast \mathcal{H} + O E)^t \wedge A\, (N \ast H + O
E) + 2 \mathcal{H}^t N^t \wedge A O E\nonumber\\
& & - 2 i {\rm Re} \mathcal{N}_{IJ} \d \chi^I \wedge (H^J - \chi^J
H)\ .
\end{eqnarray}
Here we have defined the vectors
\begin{eqnarray}
\mathcal{H} = \left( \begin{array}{c} H\\ H^I \end{array} \right)\
,\quad E = \left( \begin{array}{c} \d \phi\\
\e^{-\frac{\phi}{2}} \d \chi^I \end{array} \right)\ ,
\end{eqnarray}
and the matrices
\begin{eqnarray}
N = \e^{\frac{\phi}{2}} \left( \begin{array}{cc} \e^{\frac{\phi}{2}} & 0\\
- \chi^I & \delta^I_{\, \, J} \end{array} \right)\ ,\quad A =
\left(
\begin{array}{cc}
 \frac{1}{2} & 0\\
0 & - {\rm Im} \mathcal{N}_{IJ}\end{array} \right)\ ,
\end{eqnarray}
$O$ is a matrix as well, satisfying $O^t A O = A$. When all fields
are taken real, clearly the real part of (\ref{bbound}) is bounded
from below by
\begin{eqnarray}\label{bbo}
{\rm Re} {\cal L}^e_{T4} & \geq & \d^{4} x \, e R
+ 2 \mathcal{M}_{IJ} \ast \d X^I \wedge \d \bar{X}^J\nonumber\\
& & + 2 \mathcal{H}^t N^t \wedge A O E\ .
\end{eqnarray}
Next we take the matrix $O$ to be
\begin{eqnarray}
O_{1,2} = \pm \left( \begin{array}{cc} 1 & 0\\ 0 & \epsilon
\end{array} \right)\ ,
\end{eqnarray}
with $\epsilon = \delta^I_{\, \, J}$ in the $O_1$-case and
$\epsilon = - \delta^I_{\, \, J}$ in the $O_2$-case. The plus and
minus signs refer to the instanton and the anti-instanton
respectively.

We now consider configurations for which the square in
(\ref{bbound}) is zero (i.e that saturate the bound (\ref{bbo}) in
case all fields are taken real). It is easy to show that for
constant $X^I$ these configurations satisfy the field equations of
$\phi$, $\chi^I$ and the tensors (to the field equations of $X^I$
we come back at a later stage). Furthermore these configurations
can be shown to have vanishing energy-momentum. Therefore the
gravitational background should be flat and (\ref{bbound}) reduces
to a total derivative. In the following we need the explicit form
of this total derivative in the $O_2$-case
\begin{eqnarray}
{\cal L}^{i.}_{2} & = & - \d (\e^{\phi} H) - 2i \d
(\bar{\mathcal{N}}_{IJ}
\chi^I(H^J - \frac{1}{2} \chi^J H))\ ,\nonumber\\
{\cal L}^{a.i.}_{2} & = & + \d (\e^{\phi} H) - 2i \d (\mathcal{N}_{IJ}
\chi^I(H^J - \frac{1}{2} \chi^J H))\ ,
\end{eqnarray}
where the upper equation corresponds to the instanton and the
lower one to the anti-instanton. In the $O_1$-case we get similar
expressions.

Again we can make contact with the double-tensor multiplet theory
by taking the function $F$ to be $F(X) = - \frac{1}{4} i (X^0)^2$.
The analysis above then reduces to the analysis of \cite{TV1},
with the matrices $O_{1,2}$ corresponding to their matrices
$O_{1,2}$. The instantons related to these matrices are the
NS-fivebrane instantons discussed in section \ref{univ}.

\vspace{5mm}

Let us now consider the conditions which follow from requiring the
square in (\ref{bbound}) to vanish. Firstly, the $O_1$ matrix
gives
\begin{eqnarray}\label{o1}
\ast \mathcal{H} = \pm \left( \begin{array}{c} \d \e^{-\phi}\\
\chi^I \d \e^{-\phi} - \e^{-\phi} \d \chi^I \end{array} \right)\ .
\end{eqnarray}
These equations are very similar to the $O_1$ equations of
\cite{TV1}. Note in particular the relation $\ast H = \pm \d
\e^{-\phi}$, which is contained in both. Similarly to \cite{TV1}
we find that the finite-action-solution to (\ref{o1}) has a
harmonic $\e^{-\phi}$ and constant $\chi^I = \chi^I_0 =
\frac{Q^I}{Q}$. Here
\begin{equation}
Q^I = \int_{S^3_{\infty}} H^I\ ,\qquad Q = \int_{S^3_{\infty}} H\ ,
\end{equation}
consistent with the notation we used in our treatment of D-brane
instantons.

\vspace{3mm}

The conditions following from taking the matrix $O_2$ in
(\ref{bbound}) are
\begin{eqnarray}\label{o2}
\ast \mathcal{H} = \pm \d \left( \begin{array}{c} \e^{-\phi}\\
\e^{-\phi} \chi^I \end{array} \right)\ .
\end{eqnarray}
These equations are very similar to the $O_2$ equations of
\cite{TV1}, with once more $\ast H = \pm \d \e^{-\phi}$ contained
in both sets. The latter equation implies that $\e^{-\phi}$ is
again harmonic. The remaining equations in (\ref{o2}) tell us that
the same is true for $\e^{-\phi} \chi^I$. For single-centered
solutions this allows us to write $\chi^I$ as
\begin{equation}\label{eqchi}
\chi^I = \chi^I_1 \e^{\phi} + \chi^I_0\ ,
\end{equation}
where the $\chi^I_1$ are arbitrary constants. Note that $\chi^I_0$
is the value $\chi^I$ takes at the excised point(s). Putting
(\ref{eqchi}) back into (\ref{o2}) we find again $\chi^I_0 =
\frac{Q^I}{Q}$. Observe that the finite-action $O_1$-solution is
contained in this $O_2$-solution; we re-obtain it when we put
$\chi^I_1$ to zero. The action becomes for the single-centered
instanton
\begin{equation}\label{insaco2}
S_{\rm inst} = \frac{|Q|}{g_s^2} + i \bar{\mathcal{N}}_{IJ} \Delta \chi^I
\Delta \chi^J Q\ ,
\end{equation}
where we have (again) defined $\Delta \chi^I \equiv
\chi^I|_{\infty} - \chi^I|_{\vec{x} \rightarrow \vec{x}_o}$. For
the anti-instanton $\bar{\mathcal{N}}$ should be replaced by
$\mathcal{N}$. The equations of motion of $X^I$ are not
automatically satisfied. Requiring this gives the extra condition
that the last term in (\ref{insaco2}) should be extremized with
respect to (the constants) $X^I$. Consequently the $\chi^I_1$ and
the $X^I$ become related, unless $\mathcal{N}_{IJ} = cst$. The
latter is for example the case in the double-tensor multiplet
theory, in which we have $\mathcal{N}_{00} = - \frac{i}{2}$ (in
that case (\ref{insaco2}) can be seen to reduce to (\ref{dtns})).
The precise relations between $\chi^I_1$ and the $X^I$ depend on
the function $F(X)$. This implies that there is no general
prescription for obtaining real solutions.

{}From $\chi^I_0 = \frac{Q^I}{Q}$ it directly follows that the
charges $\hat{Q}^I \equiv Q^I - \chi^I_0 Q$ are zero. Furthermore
one can show that there are no sources in the field equations of
$\chi^I$. So there are no RR charges at all in the solutions. This
means that they can be identified as (generalized) NS-fivebrane
instantons. On the basis of what we know about NS-fivebrane
instantons in the double-tensor multiplet \cite{DTV} we expect
these solutions (or at least all single-centered ones) to preserve
half of the supersymmetry.

Let us finish our treatment of (generalized) NS-fivebrane
instantons by considering its image under the (inverse) c-map. We
find that this is a Taub-NUT geometry with $n$ (anti-)selfdual
vector fields, all of the form (\ref{grvp}). It would be
interesting to study if there are more general Euclidean BPS
solutions of this type. We leave this for further study.

\section*{Acknowledgements}

This work is partly supported by NWO grant 047017015, EU contracts
MRTN-CT-2004-005104 and MRTN-CT-2004-512194, and INTAS contract 03-51-6346.

\appendix

\section{Conventions}

\subsection{Form notation}\label{form}

Greek indices are $D$-dimensional curved indices. $a,b,...$
are $D$-dimensional flat indices, $m,n,...$ are
$D-1$-dimensional curved indices, and $i,j,...$ are $D-1$-dimensional
flat indices. The case relevant for us is of course $D=4$, but formulae
below hold for arbitrary $D$:

\begin{eqnarray}
a_{p} & \equiv & \frac{1}{p!} a_{\alpha_{1}...\alpha_{p}} \d
x^{\alpha_{1}} \wedge ... \wedge \d x^{\alpha_{p}}\ ,\nonumber\\
\d x^{\alpha_{1}} \wedge ... \wedge \d x^{\alpha_{p}} & \equiv &
e \epsilon^{\alpha_{1}...\alpha_{p}} \d^{p} x\ ,\nonumber\\
a_{p} \wedge b_{q} & \equiv & \frac{1}{p!q!}
a_{\alpha_{1}...\alpha_{p}} b_{\alpha_{p+1}...\alpha_{p+q}} \d
x^{\alpha_{1}} \wedge ...
\wedge \d x^{\alpha_{p + q}}\ ,\nonumber\\
\epsilon^{\alpha_{1}...\alpha_{p}} & \equiv &
e_{a_{1}}^{\alpha_{1}}...e_{a_{p}}^{\alpha_{p}}
\epsilon^{a_{1}...a_{p}}\ ,\nonumber\\
\epsilon^{0...p-1} & \equiv & 1\ ,\nonumber\\
\epsilon^{1...p} & \equiv & 1\ ,\nonumber\\
\ast a_{p} & \equiv & \frac{1}{(D-p)!} (\ast
a)_{\alpha_{1}...\alpha_{D-p}} \d x^{\alpha_{1}} \wedge ...
\wedge \d x^{\alpha_{D-p}}\ ,\nonumber\\
(\ast a)_{\alpha_{1}...\alpha_{D-p}} & \equiv & \frac{1}{p!}
\epsilon_{\alpha_{1}...\alpha_{D-p}}^{\; \; \; \; \; \; \; \; \;
\beta_{1}...\beta_{p}} a_{\beta_{1}...\beta_{p}}\ ,\nonumber\\
|a_{p}|^{2} & \equiv & \ast a_{p} \wedge
a_{p}\nonumber\\
& = & (-)^{s} \; e \; \frac{1}{p!} a^{\alpha_{1}...\alpha_{p}}
a_{\alpha_{1}...\alpha_{p}} \d^D x\ ,\nonumber\\
|| \alpha_{p} ||^{2} & \equiv & \ast a_{p} \wedge \bar{a}_{p}\ ,
\end{eqnarray}
with $s=1$ for Minkowskian and $s=0$ for Euclidean signature.

\subsection{Dimensional reduction}\label{dimred}

When $a^m$/ $a^e$ is a $p$-form in $D$ Minkowskian/ Euclidean
dimensions we denote the corresponding $p-1$-form in $D-1$
dimensions by the same symbol
\begin{eqnarray}
a^{m} & \equiv & \frac{1}{(p-1)!} a_{t m_{1}... m_{p-1}} \d
x^{m_{1}} \wedge ... \wedge \d x^{m_{p-1}}\ ,
\end{eqnarray}
or
\begin{eqnarray}
a^{e} & \equiv & \frac{1}{(p-1)!} a_{m_{1}... m_{p-1} \tau} \d
x^{m_{1}} \wedge ... \wedge \d x^{m_{p-1}}\ ,
\end{eqnarray}
Only in case confusion might arise we write it as
$a^m_{D-1}$/$a^e_{D-1}$.

The corresponding $p$-form in $D-1$ dimensions is denoted by
\begin{eqnarray} \mathbf{a} & \equiv &
\frac{1}{p!} a_{m_{1}... m_{p}} \d x^{m_{1}} \wedge ... \wedge \d
x^{m_{p}}\ .
\end{eqnarray}

\subsection{Wick rotation}\label{wic}

The standard Wick rotation
\begin{equation}
t  =  - i \tau\ ,
\end{equation}
defines Euclidean Lagrangians
\begin{equation}
{\cal L}_{m}  =  i {\cal L}_{e}\ .
\end{equation}
{}From their definitions it follows that
\begin{eqnarray}
e_{m} & = & e_{e}\ ,\nonumber\\
\epsilon^{t \alpha_{1}...\alpha_{D-1}} & = & (-)^{D-1}
\epsilon^{\tau \alpha_{1}...\alpha_{D-1}}\ .
\end{eqnarray}
The Wick rotation on two-forms is
\begin{equation}
B_{tm}\rightarrow i B_{\tau m}\ ,\qquad B_{mn}\rightarrow B_{mn}\ ,
\end{equation}
and similarly for one-forms.

\subsection{Integration of spherically symmetric harmonic
functions}

\begin{eqnarray}
\int_{S^{D-1}_{\infty}} \ast {\rm d} \frac{Q}{(D-2)(Vol S^{D-1})r^{D-2}}
= (-)^D Q\ .
\end{eqnarray}

\section{Electric-magnetic duality}\label{emduality}
Suppose we have a theory with a set of $n+1$ two-forms $B^I$ and
$n+1$ scalars $\chi^I$ (with $I$ running from $0$ to $n$),
possibly accompanied by other fields (collectively denoted by
$\phi$). Furthermore assume we can describe the set of two-forms
and scalars by (generalized) field strengths $C^I$ and $D^I$. The
$C^I$ are three-forms composed of the field strengths of the
two-forms with possible extra terms. The $D^I$ are one-forms
composed of the "field strengths" of the scalars with also extra
terms allowed. We then introduce the objects
\begin{eqnarray}
(\ast C_I)_{\mu} & = & \frac{1}{2} \frac{1}{\sqrt{|g|}}
\frac{\delta {\cal L}}{\delta D^{\mu I}}\ , \nonumber\\
(\ast D_I)_{\alpha \beta \gamma} & = & - \frac{1}{2}
\frac{6}{\sqrt{|g|}} \frac{\delta {\cal L}}{\delta C^{\alpha \beta
\gamma I}}\ .
\end{eqnarray}
In case the theory under consideration has only terms quadratic in
$C^I$ and/or $D^I$ it can be written as
\begin{eqnarray}\label{lrr}
{\cal L} & = & D_I \wedge C^I + D^I \wedge C_I\ .
\end{eqnarray}
We now restrict ourselves to cases where the set of equations
formed by the Bianchi identities of the generalized field
strengths and the equations of motion of $B^I$ and $\chi^I$ can be
formulated as
\begin{eqnarray}\label{bieom}
\d \left( \begin{array}{c} C^I\\ C_I \end{array} \right) & = &
\alpha ( \phi )\w \ast \left( \begin{array}{c} C^I\\ C_I
\end{array} \right) + \beta ( \phi ) \w \left( \begin{array}{c}
D^I\\ D_I \end{array} \right)\ ,\nonumber\\
\d \left( \begin{array}{c} D^I\\ D_I \end{array} \right) & = &
\gamma ( \phi )\w \ast \left( \begin{array}{c} C^I\\ C_I
\end{array} \right) + \eta ( \phi ) \w \left( \begin{array}{c}
D^I\\ D_I \end{array} \right)\ .
\end{eqnarray}
$\alpha (\phi)$ and $\beta (\phi)$ are three-forms, while $\gamma
(\phi)$ and $\eta (\phi)$ are one-forms\footnote{In section
\ref{gen} of the main text we have $\alpha = \gamma = \eta = 0$
and $\beta = i H$.}.

The set of equations (\ref{bieom}) is invariant under the
electric-magnetic duality transformations
\begin{eqnarray}\label{emdu}
\left( \begin{array}{c} C^I\\ C_I \end{array} \right)
\longrightarrow \left( \begin{array}{c} \tilde{C}^I\\ \tilde{C}_I
\end{array} \right) & = & \left( \begin{array}{cc} U^I_{\, \, \, \, J} & Z^{IJ}\\ W_{IJ} &
V_I^{\, \, \, \, J} \end{array} \right) \left( \begin{array}{c} C^J\\
C_J
\end{array} \right)\ ,\nonumber\\
\left( \begin{array}{c} D^I\\ D_I \end{array} \right)
\longrightarrow \left( \begin{array}{c} \tilde{D}^I\\ \tilde{D}_I
\end{array} \right) & = & \left( \begin{array}{cc} U^I_{\, \, \, \, J} & Z^{IJ}\\ W_{IJ} &
V_I^{\, \, \, \, J} \end{array} \right) \left( \begin{array}{c} D^J\\
D_J
\end{array} \right)\ ,
\end{eqnarray}
and $\alpha$, $\beta$, $\gamma$, and $\eta$ transforming
trivially. We call (\ref{emdu}) electric-magnetic duality
transformations as Bianchi identities and equations of motion are
rotated into each other, which is similar to the effect of
conventional electric-magnetic duality transformations working on
the field strengths and dual field strengths of one-form gauge
fields.

A dual Lagrangian is defined by
\begin{eqnarray}
(\ast \tilde{C}_I)_{\mu} & = & \frac{1}{2} \frac{1}{\sqrt{|g|}}
\frac{\delta \tilde{{\cal L}}}{\delta \tilde{D}^{\mu I}}\ ,\nonumber\\
(\ast \tilde{D}_I)_{\alpha \beta \gamma} & = & - \frac{1}{2}
\frac{6}{\sqrt{|g|}} \frac{\delta \tilde{{\cal L}}}{\delta
\tilde{C}^{\alpha \beta \gamma I}}\ .
\end{eqnarray}
In terms of the old $(C^I, C_I)$ and $(D^I, D_I)$ these equations
are
\begin{eqnarray}
\frac{1}{2} \frac{1}{\sqrt{|g|}} \frac{\delta \tilde{{\cal L}}}{\delta
D^{\mu J}} & = & (U^T W)_{JK}  \ast C^K_{\mu} + (U^T V)_J^{\, \,
\, \, K} \ast C_{\mu K}\nonumber\\
& & + \frac{\delta D_K^{\eta}}{\delta D^{\mu J}} (Z^T V)^{KL} \ast
C_{\eta L} - \frac{1}{6} \frac{\delta C^{\alpha \beta
\gamma}_K}{\delta
D^{\mu J}} (Z^T V)^{KL} \ast D_{\alpha \beta \gamma L}\nonumber\\
& &  + \frac{\delta D_K^{\eta}}{\delta D^{\mu J}} (Z^T W)^K_{\, \,
\, \, L} \ast C_{\eta}^L - \frac{1}{6} \frac{\delta C^{\alpha
\beta
\gamma}_K}{\delta D^{\mu J}} (Z^T W)^K_{\, \, \, \, L} \ast D_{\alpha \beta \gamma}^L\ ,\nonumber\\
\frac{1}{2} \frac{1}{\sqrt{|g|}} \frac{\delta \tilde{{\cal L}}}{\delta
C^{\mu \rho \sigma J}} & = & - \frac{1}{6} (U^T W)_{JK} \ast
D^K_{\mu \rho \sigma} - \frac{1}{6} (U^T V)_J^{\, \, \, \, K} \ast
D_{\mu \rho \sigma K}\nonumber\\ & & - \frac{1}{6} \frac{\delta
C_K^{\alpha \beta \gamma}}{\delta C^{\mu \rho \sigma J}} (Z^T
V)^{KL} \ast D_{\alpha \beta \gamma L} + \frac{\delta
D^{\eta}_K}{\delta C^{\mu \rho \sigma J}} (Z^T V)^{KL} \ast C_{\eta L}\nonumber\\
& &  - \frac{1}{6} \frac{\delta C_K^{\alpha \beta \gamma}}{\delta
C^{\mu \rho \sigma J}} (Z^T W)^K_{\, \, \, \, L} \ast D_{\alpha
\beta \gamma}^L + \frac{\delta D^{\eta}_K}{\delta C^{\mu \rho
\sigma J}} (Z^T W)^K_{\, \, \, \, L} \ast C_{\eta}^L\ .
\end{eqnarray}
As it turns out this set of equations can only be solved
consistently in case the transformation matrix in (\ref{emdu})
belongs to $Sp(2n +2, \mathbb{R})$ when all fields are real, or
otherwise a complexified version thereof. Furthermore it is
important to note that generically we get ${\cal L} (\tilde{C}^I,
\tilde{D}^I) \neq {\cal L} (C^I, D^I)$, i.e. the Lagrangian does not
transform as a function. For purely quadratic theories the dual
Lagrangian becomes
\begin{eqnarray}\label{lrrt}
\tilde{{\cal L}} & = & \tilde{D}_I \wedge \tilde{C}^I + \tilde{D}^I
\wedge \tilde{C}_I\ .
\end{eqnarray}


\end{document}